\documentclass[nofootinbib,a4paper,aps,prd,10pt,superscriptaddress,reprint,showkeys,showpacs]{revtex4-1}
\usepackage{graphics}
\usepackage{epsfig} 
\usepackage{amsmath}
\usepackage{multirow}
\usepackage{rotating}
\usepackage{float}
\usepackage{hyperref}
\usepackage{color}
\usepackage{soul}
\hypersetup{%
    ,urlcolor=blue
    ,citecolor=blue
    ,linkcolor=blue
    }

\graphicspath{{./figures/}}

\renewcommand{\arraystretch}{1.5}

\input epsf

\newcommand{\be}{\begin{equation}}
\newcommand{\ee}{\end{equation}}
\newcommand{\bea}{\begin{eqnarray}}
\newcommand{\eea}{\end{eqnarray}}
\def\pslash{{\cal P}{\hbox{\kern-6pt $\slash$}}}

\def\aap{A\&A}
\def\apj{ApJ}
\def\apjl{ApjL}
\def\apjs{ApJS}
\def\araa{ARA\&A}
\def\baas{Bulletin of the American Astronomical Society}
\def\jcap{JCAP}
\def\mnras{MNRAS}
\def\nar{New Astronomy Reviews}
\def\physrep{Physics Reports}

\long\def\comment#1{}

\newcommand{\rmd}{\mathrm{d}}

\begin{document}
\title{Spherical collapse in generalized dark matter models}

\author{F. Pace}
\email{francesco.pace@manchester.ac.uk}
\affiliation{Jodrell Bank Centre for Astrophysics, School of Natural Sciences, Department of Physics and Astronomy, The University of Manchester, Manchester, M13 9PL, United Kingdom}

\author{Z. Sakr}
\email[Corresponding author: ]{ziad.sakr@net.usj.edu.lb}
\affiliation{Physics Department, University of Saint Joseph, Lebanon}

\author{I. Tutusaus}
\email{tutusaus@ice.csic.es}
\affiliation{Institute of Space Sciences (ICE, CSIC), Campus UAB, Carrer de Can Magrans, s/n, 08193 Barcelona, Spain and Institut d'Estudis Espacials de Catalunya (IEEC), Carrer Gran Capit\`a 2-4, 08193 Barcelona, Spain}

\label{firstpage}

\date{\footnotesize{Received 30 December 2019; accepted 14 July 2020; published 10 August 2020}}

\begin{abstract}
The influence of considering a generalized dark matter (GDM) model, which allows for a non-pressure-less dark matter and a nonvanishing sound speed in the nonlinear spherical collapse model is discussed for the Einstein-de Sitter-like and $\Lambda$GDM models. 
By assuming that the vacuum component responsible for the accelerated expansion of the Universe is not clustering and therefore behaving similarly to the cosmological constant $\Lambda$, we show how the change in the GDM characteristic parameters affects the linear density threshold for collapse of the nonrelativistic component ($\delta_{\rm c}$) and its virial overdensity ($\Delta_{\rm V}$). We found that a positive GDM equation of state parameter, $w_{\rm gdm}$, is responsible for lower values of $\delta_{\rm c}$ as compared to the standard spherical collapse model and that this effect is much stronger than the one induced by a change in the GDM sound speed, $c^2_{\rm s, gdm}$. We also found that $\Delta_{\rm V}$ is only slightly affected and mostly sensitive to $w_{\rm gdm}$. These effects could be relatively enhanced for lower values of the matter density. We found that the effects of the additional physics on $\delta_{\rm c}$ and $\Delta_{\rm V}$, when translated to nonlinear observables such as the halo mass function, induce an overall deviation of about 40\% with respect to the standard $\Lambda$CDM model at late times for high mass objects. However, within the current constraints for $c^2_{\rm s, gdm}$ and $w_{\rm gdm}$, we found that these changes are the consequence of properly taking into account the correct linear matter power spectrum for the GDM model while the effects coming from modifications in the spherical collapse model remain negligible. Using a phenomenologically motivated approach, we also study the nonlinear matter power spectrum and found that the additional properties of the dark matter component lead, in general, to a strong suppression of the nonlinear power spectrum with respect to the corresponding $\Lambda$CDM one. Finally, as a practical example, we compare $\Lambda$GDM and $\Lambda$CDM using galaxy cluster abundance measurements, and found that these small scale probes will allow us to put more stringent constraints on the nature of dark matter.
\end{abstract}

\pacs{95.35.+d, 98.80.-k, 98.80.Es}

\keywords{Cosmology; Generalized Dark Matter; Mass function; Spherical collapse model; Non-linear matter power spectrum}

\maketitle

\section{Introduction}
Nowadays, most of the cosmological data suggests a cosmic expansion history with a flat geometry and some sort of dark energy, usually in the form of the cosmological constant $\Lambda$, in order to explain the recent accelerating expansion of the Universe. Assuming that large scale structure formed thanks to the gravitational interaction of the cold dark matter (CDM) component, the resulting standard model of cosmology is then dubbed $\Lambda$CDM (see \cite{Frieman2008,Spergel2015} for a review). In this model, the cosmological constant is a fluid with constant equation of state $w=-1$ and energy density $\rho_{\Lambda}$, both constant in time and space, that is usually associated to the vacuum energy density. The CDM component is instead described as a nonrelativistic fluid whose influence is only gravitational. Together, the cosmological constant and the CDM amount to approximately 95\% of the total energy budget, with the remaining 5\% in the form of baryons and a negligible amount, today, of relativistic particles (photons and neutrinos) \citep{Planck2018_VI}.
 
However, with the advent of stage IV surveys like DESI \citep{DESI2016}, Euclid \citep{Laureijs2011}, LSST \citep{Marshall2017} , WFIRST \citep{Dore2019}, and the SKA \citep{SKA2018} providing high accuracy data especially on small scales, one of the most challenging problems is to understand the role played by the different cosmic components in the nonlinear regime of gravitational clustering. This aspect could be tackled through different approaches, among which we recall the halo model \citep{Cooray2002}, where one of the issues is to understand the interplay of different possible physical effects that contribute to determine the properties of virialized halos. One of the powerful tools to study the nonlinear evolution of perturbations and formation of haloes is given by the popular spherical collapse model (SCM), introduced in a seminal paper by \cite{Gunn1972} to deal with a system made only of CDM. This model has been later extended and applied to study the evolution of density perturbations and structure formation in the presence of dark energy, both homogeneous \citep{Wang1998,Horellou2005,Pace2010,Pace2012,Pace2014,Pace2019b} and clustering \cite{Batista2013,NazariPooya2016,Pace2017a}. In this work, we investigate further the nonlinear evolution of matter perturbations by focusing on the generalized dark matter (GDM) model, which considers the dark matter fluid augmented by positive pressure, parametrized by a background equation of state $w_{\rm gdm}$ and a nonvanishing sound speed $c_{\rm s, gdm}^2$ \citep{Hu1998}. Note that the original GDM description also accounted for negative pressure and a nonvanishing viscosity $c_{\rm vis, gdm}^2$. The effects of this modeling on the expansion and linear perturbations have been recently studied in \cite{Kopp2016,Kunz2016,Thomas2016,Kopp2018,Tutusaus2018,Thomas2019}. Here we decided not to include the contribution of the viscosity $c_{\rm vis, gdm}^2$ since at the linear level it is degenerate with the sound speed $c_{\rm s, gdm}^2$. This can be seen considering the scale at which the potential decays at a conformal time $\eta$ \citep{Kopp2016,Thomas2016}:
 \begin{equation}
  k_{\rm d}^{-1}(\eta) = \eta\sqrt{c_{\rm s, gdm}^2+\frac{8}{15}c_{\rm vis, gdm}^2}\,.
 \end{equation}
 On scales larger than $k_{\rm d}^{-1}$ one cannot distinguish the effect of these parameters separately. The viscosity, in addition, causes a dump in the oscillations, but the cutoff is a bigger effect. This degeneracy would be broken in the time-dependent case, however analysis \cite{Kopp2018} showed that current data are not good enough to spot any evidence of a time evolution for any of the GDM parameters. Therefore, we only consider constant background equation-of-state parameter and sound speed.

However, the standard SCM needs to be appropriately modified to have a recipe for the GDM to be able to explore the small scales that next stage surveys will probe and extract the maximum information from them. A recent approach, focused on developing a halo-model-based approach for nonlinear corrections for the GDM matter power spectrum, considered a scaling of the SCM \cite{Thomas2019}. Here, instead, we focus on developing the SCM for GDM by considering the evolution of matter perturbations within the GDM framework (taking into account the effects of $w_{\rm gdm}$ and $c_{\rm s,gdm}^2$), and use a simple phenomenological approach to address the nonlinear matter power spectrum. We restrict our analysis to an Einstein-de-Sitter-like (EdS) model where $\Omega_{\mathrm{m}}=1$ and $\Omega_{\Lambda}=0.0$, and a flat $\Lambda$CDM cosmology. 
For the $\Lambda$CDM model, we assume the following cosmological parameters: $\Omega_{\rm m}=0.3$, $\Omega_{\Lambda}=0.7$ and $h=0.7$. In particular, we discuss how the linear overdensity threshold for collapse ($\delta_{\rm c}$) and the virial overdensity ($\Delta_{\rm V}$) change while changing the properties of the dark matter component.

The paper is organized as follows: in Sec.~\ref{sect:gdmmod} we give a brief description of the spherical collapse model for generalized dark matter and derive the appropriate equations describing the evolution of nonlinear perturbations, by specializing on the virial overdensity $\Delta_{\rm V}$ and the linearly extrapolated overdensity $\delta_{\rm c}$. In Sec.~\ref{sect:resuL} we present our findings, studying the evolution of the main parameters of the spherical collapse model as a function of the equation of state $w_{\rm gdm}$ and effective sound speed $c^2_{\rm s, gdm}$ of the matter component and translate our results into observable quantities such as the mass function. As a practical application, we compute the goodness of fit of the $\Lambda$GDM model, for some specific values of $w_{\rm gdm}$ and $c_{\rm s,gdm}^2$, using cluster counts from real data. Finally, we discuss the evolution of the nonlinear matter power spectrum based on a phenomenologically motivated approach and conclude in Sec.~\ref{sect:concL}.

\section{The GDM model}\label{sect:gdmmod}
In this work we assume that dark matter only interacts gravitationally with the other components and all fluids satisfy the standard continuity equation $\nabla_{\nu}T_i^{\mu\nu}=0$, where $T_i^{\mu\nu}$ is the stress-energy tensor and for a perfect fluid reads as
\begin{equation}
  T_i^{\mu\nu}=(\rho_i c^2+P_i)u^{\mu}u^{\nu}+P_ig^{\mu\nu}\,,
\end{equation}
where $\rho_i$, $P_i$ and $u_i$ are the density, the pressure, and the four-velocity of each fluid, respectively, and $g^{\mu\nu}$ the metric.

Contracting the continuity equation once with $u_{\mu}$ and once with the projection operator $h_{\mu\alpha}=g_{\mu\alpha}+u_{\mu}u_{\alpha}$, one obtains the relativistic expressions for the continuity and the Euler equations, respectively:
\begin{eqnarray}
  \frac{\partial\rho_i}{\partial t} + \nabla_{\vec{r}}\cdot(\rho_i\vec{v}_i) + 
  \frac{P_i}{c^2}\nabla_{\vec{r}}\cdot\vec{v}_i = 0\,, \label{equ:cnpert} \\
  \frac{\partial\vec{v}_i}{\partial t} +  (\vec{v}_i\cdot\nabla_{\vec{r}})\vec{v}_i+
  \nabla_{\vec{r}}\Phi+\frac{\nabla_{\vec{r}}P_i}{\rho_i+P_i/c^2}=0\,. \label{equ:enpert}
\end{eqnarray}
Here $\vec{v}_i$ is the three-dimensional velocity of each species, $\Phi$ the Newtonian gravitational potential and $\vec{r}$ denotes physical coordinates.

The $00$-component of Einstein's field equations gives the relativistic Poisson equation
\begin{equation}\label{equ:pnpert}
 \nabla_{\vec{r}}^2\Phi = 4\pi G\sum_k\left(\rho_k+\frac{3P_k}{c^2}\right)\,,
\end{equation}
where the potential is sourced by all the fluid components and $\rho_k$ and $P_k$ are the total density and pressure of each fluid. We define, in fact $\rho_k=\bar{\rho}_k+\delta\rho_k$ and $P_k=\bar{P}_k+\delta P_k$, where overbarred quantities represent the background.

The background continuity equation for the fluid $i$ is
\begin{equation}
 \dot{\bar{\rho}}_i + 3H\left(\bar{\rho}_i+\frac{\bar{P}_i}{c^2}\right)=0\,,
\end{equation}
where $\bar{\rho}_i=\tfrac{3H^2\Omega_{i}}{8\pi G}$ and $\Omega_{i}$ is the fluid density parameter. To solve the previous expression, it is necessary to specify a relation between pressure and density. This is usually done by introducing the background equation-of-state parameter $w_i=\bar{P}_i/(\bar{\rho}_ic^2)$, so that one solves the equation $\dot{\bar{\rho}}_i + 3H(1 + w_i)\bar{\rho}_i = 0$, once the time dependency of $w_i$ is provided.

To study the perturbations, we introduce comoving coordinates $\vec{x}=\vec{r}/a$, with $a$ the scale factor, and define
\begin{align}
 \rho_i(\vec{x},t) = &\, \bar{\rho}_i(1+\delta_i(\vec{x},t))\,, \label{equ:rpert} \\
 P_i(\vec{x},t) = &\, \bar{P}_i + \delta P_i\,, \label{equ:ppert}\\
 \Phi(\vec{x},t) = &\, {\Phi_0}(\vec{x},t)+\phi(\vec{x},t)\,, \label{equ:fpert}\\
 \vec{v}_i(\vec{x},t) = &\, a[H(a)\vec{x}+\vec{u}_i(\vec{x},t)]\,, \label{equ:vpert}
\end{align}
where $H(a)$ is the Hubble function and $\vec{u}(\vec{x},t)$ the comoving peculiar velocity. We relate pressure perturbations to density perturbations by introducing the effective sound speed $c_{{\rm s},i}^2=\delta P_i/(\delta\rho_ic^2)$. In a standard cold dark matter model, $c_{{\rm s},i}^2=0$, as there are no pressure perturbations.

Inserting Eqs.~(\ref{equ:rpert})--(\ref{equ:vpert}) into Eqs.~(\ref{equ:cnpert})--(\ref{equ:pnpert}), and taking into account the background equations, we derive the following equations for the perturbed quantities:
\begin{align}
& \dot{\delta}_i + 3H\left(c_{{\rm s},i}^{2}-w_i\right)\delta_i  = 
 -\left[1+w_i+\left(1+c_{{\rm s},i}^{2}\right)\delta_i\right] 
\vec{\nabla}\cdot\vec{u}_i\,, \label{equ:cont-pert2}\\
& \dot{\vec{u}}_i+2H\vec{u}_i+(\vec{u}_i\cdot\vec{\nabla})\vec{u}_i+\frac{\vec{\nabla}\phi}{a^2} = 0\,, \label{equ:euler-pert2}\\
& \nabla^{2}\phi = 4\pi Ga^2\sum_k\bar{\rho}_{k}\left(1+3c_{{\rm s}, k}^{2}\right)\delta_k\,. \label{equ:poisson-pert2}
\end{align}
Note that, as commonly done, we assumed a top-hat profile for the density perturbations. This leads to $\vec{\nabla}\delta_i=0$, which considerably simplifies the equations. 
In addition, both $w_i$ and $c_{{\rm s},i}^{2}$ are functions of time only. While this is justified for the equation of state, it is a simple approximation for the sound speed, but nevertheless in agreement with current literature \cite{Kopp2016}.

The previous sets of equations allow us to study the evolution of the linearly extrapolated overdensity $\delta_{\rm c}$, which represents an important ingredient for the mass function, a tool used to infer the effects of dark energy and modified gravity on some observables like cluster abundance. In this work, we will follow a similar line of thinking, but use the mass function to test properties of the dark matter component, rather than the gravitational sector.

In full generality, to derive the equation of motion of the (generalized) dark matter component, one takes the time derivative of Eq.~(\ref{equ:cont-pert2}) and substitutes in it the divergence of Eqs.~(\ref{equ:euler-pert2}) and (\ref{equ:poisson-pert2}). Nevertheless, when doing so, the final expression becomes very complicated as both the equation of state and the effective sound speed can be time dependent. This expression will give very little insight to understand the physics of the problem. We will, therefore, first derive the full equation by defining additional coefficients which will help to write the final result in a rather compact form and, subsequently, we will specialize it to the simpler case where $c_{\rm s}^2=0$, but $w\neq0$. This will correspond to the case where dark matter fully clusters.

Following \cite{Abramo2007}, we define the following quantities:
\begin{equation*}
 A_i \equiv 3H\left(c_{{\rm s},i}^{2} - w_i\right)\delta_i\,, \quad 
 B_i \equiv 1+w_i + \left(1+c_{{\rm s},i}^{2}\right)\delta_i\,,
\end{equation*}
so that Eq.~(\ref{equ:cont-pert2}) can be written as
\begin{equation}\label{equ:cont_sh}
 \dot{\delta}_i + A_i + B_i\theta_i = 0\,,
\end{equation}
where $\theta_i \equiv \vec{\nabla}\cdot\vec{u}_i$.

At the same time, the divergence of Eq.~(\ref{equ:euler-pert2}) can be written as
\begin{equation}\label{equ:euler_sh}
 \dot{\theta}_i+2H\theta_i+\frac{1}{3}\theta_i^2+\frac{\nabla^2\phi}{a^2}=0\,,
\end{equation}
where spherical symmetry is assumed.

Taking the time derivative of (\ref{equ:cont_sh}) and using Eq.~(\ref{equ:euler_sh}) to replace $\dot{\theta}_i$ and Eq.~(\ref{equ:cont_sh}) for $\theta_i$, we finally get
\begin{equation}\label{equ:newt_sc}
 \begin{split}
  \ddot{\delta}_i + \dot{A}_i + \left(2H-\frac{\dot{B}_i}{B_i}\right)\left(A_i+\dot{\delta}_i\right) - \\
  \frac{1}{3}\frac{\left(\dot{\delta}_i+A_i\right)^2}{B_i} - \frac{B_i}{a^2}\nabla^2\phi = 0 \,.
 \end{split}
\end{equation}

For $c_{\rm s}^2=w=0$, $A_i=0$ and $B_i=1+\delta_i$, leading to the standard equation describing matter perturbations in the presence of the cosmological constant or smooth dark energy:
\begin{equation}
 \ddot{\delta}_i + 2H\dot{\delta}_i - \frac{4}{3}\frac{\dot{\delta}_i^2}{1+\delta_i} - 
 4\pi G\sum_k\bar{\rho}_k\delta_k = 0\,.
\end{equation}

Note that here the sum over the perturbed species is done for baryons (considered to be a pressureless fluid) and (generalized) dark matter. Therefore, we need to solve two differential equations of motion for the perturbations, one for baryons and one for generalized dark matter. Nevertheless, since baryons are subdominant at all times, considering only the GDM component would not alter our conclusions.

Let us now consider a specific case where $c_{{\rm s},i}^2=w_i$ to grasp more understanding of the evolution of matter perturbations. The previously defined coefficients simplify to $A_i=0$ and $B_i=(1+w_i)(1+\delta_i)$, and perturbations are adiabatic, with $P_{\rm gdm}=w_{\rm gdm}c^2\rho_{\rm gdm}$ also at the perturbative level. Similarly to what was shown in \cite{Pace2010} for homogeneous dark energy models, the equation of motion now reads
\begin{equation}\label{equ:wnldeq}
 \begin{split}
  \ddot{\delta} + \left(2H-\frac{\dot{w}_{\rm gdm}}{1+\rm w_{gdm}} \right)\dot{\delta} - 
  \frac{1}{3}\frac{4+3w_{\rm gdm}}{1+w_{\rm gdm}}\frac{\dot{\delta}^2}{1+\delta}\\
  -(1+w_{\rm gdm})(1+\delta) \frac{\nabla^{2}\phi}{a^{2}}=0\,,
 \end{split}
\end{equation}
where, for simplicity, from now on, we drop the index $i$ and consider only the expressions for generalized dark matter.

These expressions show that the nonlinear dynamics of matter perturbations can be heavily affected by the presence of a background equation-of-state parameter $w_{\rm gdm}$ and therefore we expect its value to be severely constrained. Similar conclusions can be reached for the effective sound speed $c_{\rm s}^2$, as a value different from zero defines a sound horizon scale associated to perturbations which generally implies that the fluid is not fully clustering.

Having derived the expressions for the nonlinear evolution of matter density perturbations, we stress that, despite we consider constant $w_{\rm gdm}$ and sound speed $c_{\rm s, gdm}^2$, the formalism does not rely on this assumption and can be used without modifications also in the case of time dependence.

To determine the virial overdensity $\Delta_{\mathrm{V}}$, we assume energy conservation during the collapse. This condition leads to a relation between the potential and kinetic energy of the collapsing sphere at turn-around and virialization time \citep{Lahav1991}:
\begin{equation}\label{equ:eq-virial}
 U_{\rm gdm,ta} + U_{\rm \Lambda,ta}  = U_{\rm gdm,vir} + T_{\rm gdm,vir} + U_{\Lambda,{\rm vir}} + T_{\Lambda,{\rm vir}}\,,
\end{equation}
where $U$ and $T$ are the potential and kinetic energy, respectively, of the GDM and dark energy $\Lambda$ component. The subscripts ${\rm ta}$ and ${\rm vir}$ refer to turn-around and virialization, respectively. For simplicity, we will assume the dark energy component to be in the form of a cosmological constant, but our results can be easily extended to more general models.

The potential energy for a fluid endowed with pressure as the GDM is $U_{\rm gdm} = -\tfrac{3}{5}\left(1+3w_{\rm gdm}\right)\tfrac{GM^2}{R}$ and for the cosmological constant is $U_{\Lambda} = \frac{4\pi}{5}GM\rho_{\Lambda} R^2$, where $M$ and $R$ are the mass and the radius of the spherical perturbation, respectively. For a system with the potential energy of the form $U\propto R^n$, the kinetic energy will be $T=nU/2$ \citep{Mechanics1969}. Then, according to the virial theorem, we find
\begin{equation}
 U_{\rm gdm,ta} + U_{\Lambda,{\rm ta}}  = \frac{1}{2} U_{\rm gdm,vir} + 2 U_{\Lambda,{\rm vir}}\,.
\end{equation}

Defining $\theta = \tfrac{\rho_\Lambda}{\rho_{\rm gdm}}$ and $\eta=\tfrac{r_{\rm vir}}{r_{\rm ta}}$ as in \cite{Horellou2005,Iliev2001}, we find a cubic equation describing the evolution of $\eta$
\begin{equation}\label{equ:cubiceq}
 \theta\eta^3 +\left(1+\frac{\theta}{2}\right)\eta - 1/2= 0 \,,
\end{equation}
where we used
\begin{equation}
 \left(\frac{\bar{\rho}_{\rm X, eff}}{\bar{\rho}}\right)_{\rm vir}=
 \theta\eta^3\left(\frac{a_{\rm vir}}{a_{\rm ta}}\right)^{-3(1+w_{\Lambda})} \,.
\end{equation}
In the previous expression, $w_{\Lambda}=-1$ and $\bar{\rho}_{\rm X, eff}=\bar{\rho}_{\rm X}+3\bar{P}_{\rm X}/c^2$.

Solving for $\eta$, the virial overdensity at collapse redshift $z_{\rm c}$ is
\begin{equation}\label{equ:deltavir}
 \Delta_{\rm V}(z_{\rm c}) = \frac{\rho_{\rm vir}}{\bar{\rho}_{\rm vir}} = \eta^{-3} \left.\frac{\rho_{\rm cluster}}{\bar{\rho}}\right|_{\rm ta} \left(\frac{1 + z_{\rm ta}}{1 + z_{\rm coll}} \right)^{3}\,, 
\end{equation}
where $\rho_{\rm cluster}=\bar{\rho}(1+\delta)$ is the total density of the perturbation. Since the constrained values for $w_{\rm gdm}\ll 1$ and $c_{\rm s, gdm}^2\ll1$ \citep{Thomas2016,Thomas2019,Tutusaus2018}, we assumed, for simplicity, that matter scales as in the standard CDM model.

\section{Results}\label{sect:resuL}
In this section we present some results for the spherical collapse model for the generalized dark matter models previously discussed, taking into account the effects of both the background equation-of-state parameter $w_{\rm gdm}$ and the effective sound speed $c_{\rm s, gdm}^2$. We concentrate on the linear overdensity parameter $\delta_{\mathrm{c}}$ and the virial overdensity $\Delta_{\mathrm{V}}$. These quantities have both their own theoretical importance: the linear overdensity parameter is a key ingredient for the halo mass function, while the virial overdensity is a measure of how dense cosmic structures are and ultimately, in first approximation, assuming spherical symmetry, it gives a measure of their radius knowing their mass. Whilst quite often the halo mass function is evaluated under the approximation that $\delta_{\rm c}\approx\delta_{\rm c}^{\rm EdS}$, it is necessary, in our opinion, in an era of precision cosmology where data become progressively more accurate, to perform an exact and detailed analysis to avoid introducing artificial biases in the study of the mass function which will hamper a proper comparison between analytical predictions and future observational data and lead to erroneous conclusions. The virial overdensity is indirectly related to the mass function and it can be used to determine the weak-lensing peaks, as discussed in \cite{Pace2019b}, where the authors studied the effect of different virialization recipes and were able to show that one of the recipes proposed in the literature provides results which are at odds with current numerical and observational results. In this work, we will not pursue this specific analysis, as our analysis would need to be validated with $N$-body simulations, but we will, nevertheless, comment upon it.

\subsection{Evolution of the spherical collapse parameters}
To evaluate the evolution of the spherical collapse parameters, we follow \cite{Pace2010,Pace2017a} and we look for an initial overdensity $\delta_{\rm ini}$ such that the nonlinear equation (\ref{equ:wnldeq}), in the general case where $w_{\rm gdm}\neq c_{\rm s, gdm}^2$ and both not null, diverges at the chosen collapse time. This same value is then used as an initial condition of the linearized version of (\ref{equ:wnldeq}), which describes the evolution of $\delta_{\mathrm{c}}$. The value of $\Delta_{\mathrm{V}}$, instead, simply follows by evaluating $\eta$ as explained in \cite{Horellou2005}, which, as said above, is an approximation to the true behavior of the GDM, but due to the strong constraints, this does not introduce a significant bias.

\begin{table*}
 \begin{center}
  \renewcommand*{\arraystretch}{1.1}
  \caption{\label{table_linear_constraints} Constraints on the GDM parameters using linear theory and different combinations of cosmological probes (CMB, CMB lensing, and BAO) from \citet{Thomas2019}. The values provided correspond to the 95\% confidence regions.}
  \begin{ruledtabular}
   \begin{tabular}{cccc}
    & CMB & CMB + lensing & CMB + lensing + BAO \\
    \hline
    $10^2 \times w_{\rm gdm}$ & $-0.040^{+0.473}_{-0.468}$ & $0.066^{+0.434}_{-0.427}$ & $0.074^{+0.111}_{-0.110}$ \\
    \hline
    $10^6 \times c_{\rm s,gdm}^2$ & $<3.31$ & $<1.92$ & $<1.91$ \\
    \hline
    $10^6 \times c_{\rm vis, gdm}^2$ & $<5.70$ & $<3.27$ & $<3.30$ \\
   \end{tabular}
  \end{ruledtabular}
 \end{center}
\end{table*}

\begin{figure}[!t]
\centering
\includegraphics[width=\hsize]{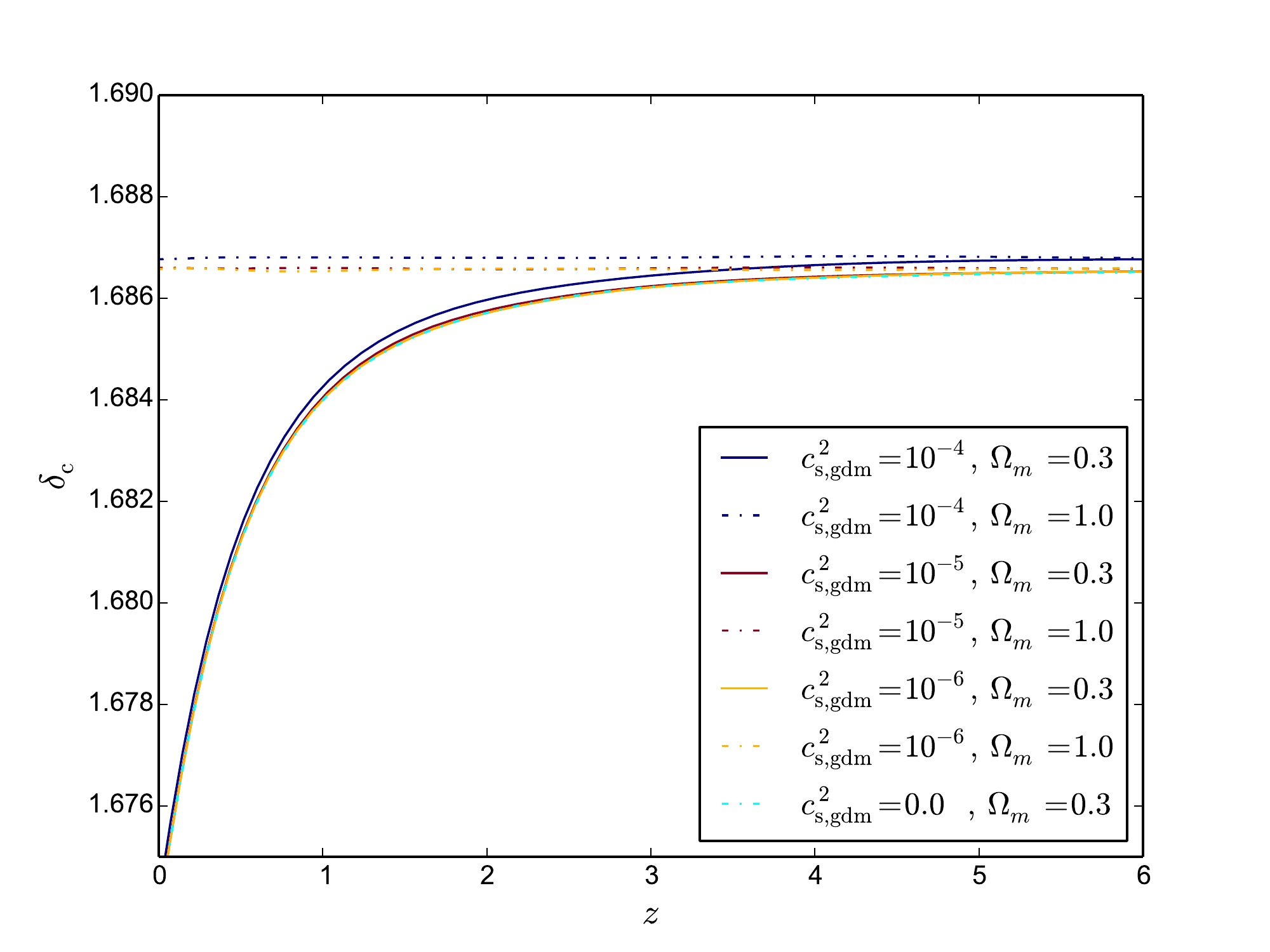}
\caption{The linear critical density contrast $\delta_{\rm c}$ as a function of the collapse redshift $z$ for different values of the generalized dark matter sound speed $c^2_{\rm s, gdm}$ for an EdSGDM and flat $\Lambda$GDM cosmology. For reference, we consider also the standard $\Lambda$CDM model, where $c_{\rm s, gdm}^2=0$.}
\label{fig:csfree}
\end{figure}

In Fig.~\ref{fig:csfree} we show the evolution of the critical overdensity $\delta_{\rm c}$ as a function of redshift $z$ assuming Einstein-de Sitter-like (EdSGDM) and flat $\Lambda$GDM as cosmological models. 
Different curves refer to different values of the effective sound speed, while keeping $w_{\rm gdm}=0$ as for the standard cold dark matter model. This setup allows us to study the effect of the modified clustering properties of dark matter. Note that for stability reasons, $c_{\rm s, gdm}^2>0$. For comparison, in cyan, we also show the evolution of the reference $\Lambda$CDM cosmology. The values chosen for the effective sound speed are motivated by the constraints obtained studying the evolution of linear perturbations \citep{Kunz2016,Tutusaus2018,Thomas2019}. For completeness, in Table~\ref{table_linear_constraints} we provide the constraints obtained in  \cite{Thomas2019}.

As expected, all the models asymptotically approach the EdS limit at high redshifts, regardless of the sound speed value. Differences for $\delta_{\rm c}$ between the $\Lambda$GDM and the standard $\Lambda$CDM model are absolutely negligible, and likely due to numeric, except for high values of the sound speed, well above the linear constraints limits, i.e., $c_{\rm s, gdm}^2\sim 10^{-4}$. This shows that to modify the evolution of $\delta_{\rm c}$, relatively high values of the sound speed are required.

As the sound speed $c_{\rm s, gdm}^2$ influences how much the fluid collapses, we can understand the dependence of $\delta_{\rm c}$ if we vary this parameter. As the sound speed increases, a higher $\delta_{\rm c}$ is needed, because there is an additional pressure effect that resists the collapse and opposes structure formation. We remind the reader that a higher value of the sound speed implies a smoother component.

We do not show the effect of the sound speed on the virial overdensity $\Delta_{\rm V}$ as this parameter is not directly included in its definition, but it enters in it through the nonlinear evolution of matter perturbations.

\begin{figure}[!t]
\centering
\includegraphics[width=\hsize]{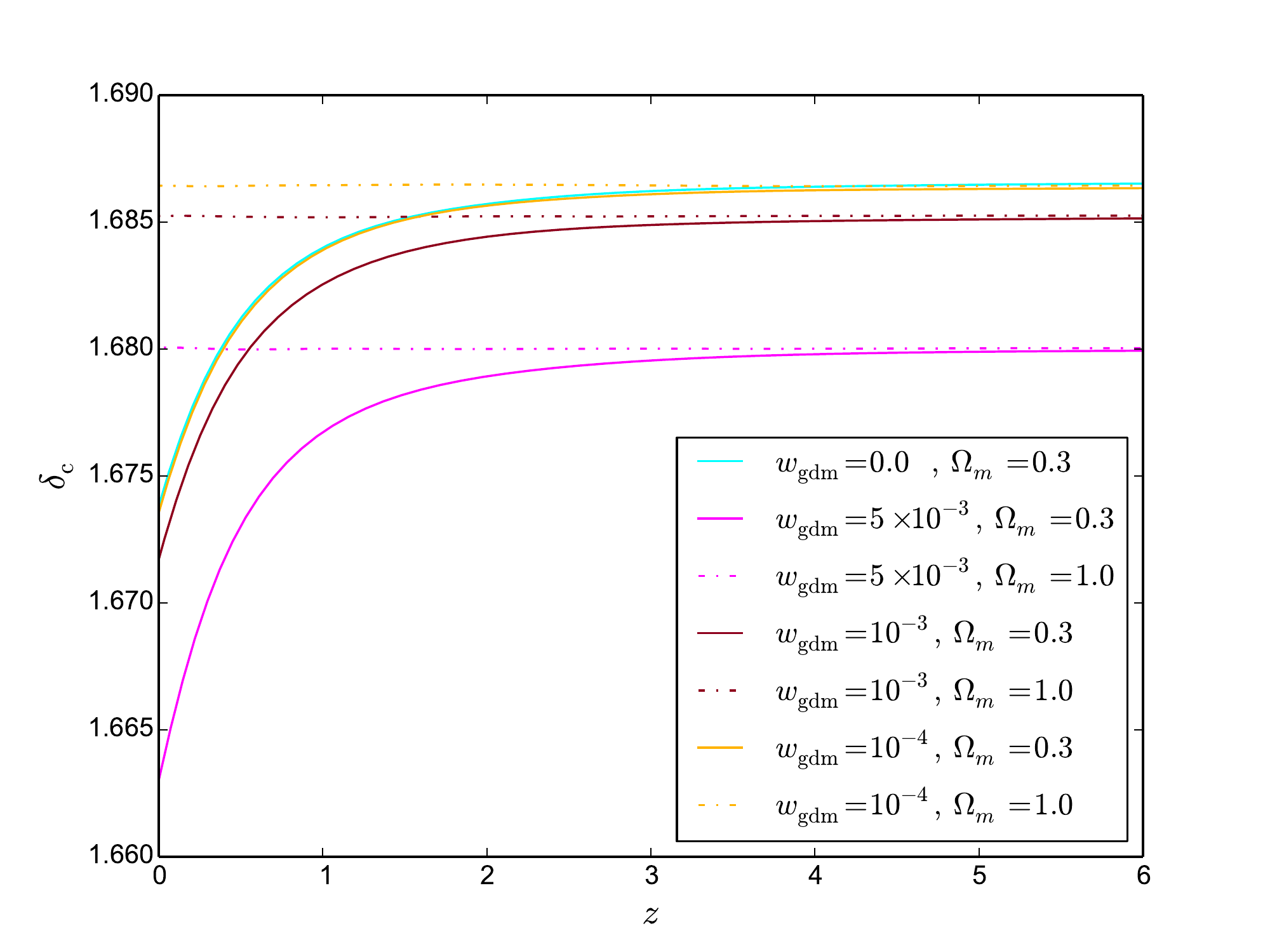}
\includegraphics[width=\hsize]{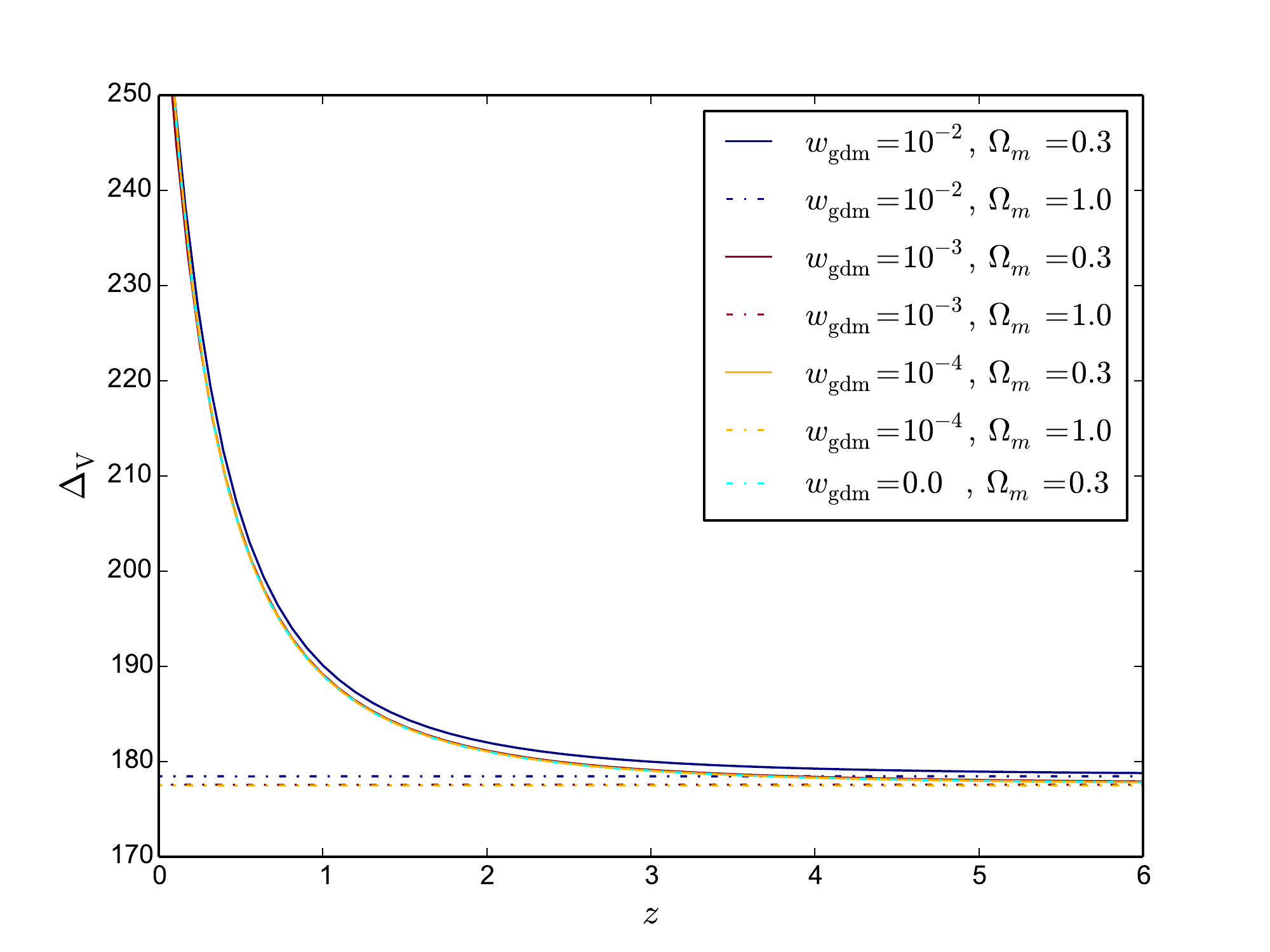}
\caption{Top (bottom) panel: The linear critical density contrast $\delta_{\rm c}$ (virial overdensity $\Delta_{\rm V}$) as a function of the collapse redshift $z$ for different values of the generalized dark matter equation of state parameter $w_{\rm gdm}$ for an EdSGDM and flat $\Lambda$GDM cosmology.}
\label{fig:wgfree}
\end{figure}

In Fig.~\ref{fig:wgfree}, we present the evolution of $\delta_{\rm c}$ (top panel) and $\Delta_{\rm V}$ (bottom panel) as a function of redshift $z$ for different values of the generalized dark matter equation-of-state parameter $w_{\rm gdm}$, while setting $c_{\rm s, gdm}^2=0$. It is immediately clear that $w_{\rm gdm}$ has a much stronger effect than that induced by the sound speed and it becomes more pronounced for $\delta_{\rm c}$ than $\Delta_{\rm V}$ for values around $w_{\rm gdm}\sim 10^{-3}$ where we also checked that the relative difference of $\delta_{\rm c}$ with the $\Lambda$CDM result is few times higher than that of $\Delta_{\rm V}$ even if we go to $w_{\rm gdm}\sim 10^{-2}$ for the latter.

To see why the equation-of-state parameter $w_{\rm gdm}$ has a stronger effect than $c_{\rm s, gdm}^2$, we remind the reader that when $w_{\rm gdm}\neq0$, the background expansion history is modified and pressure effects are not negligible even at early times, while $c_{\rm s, gdm}^2$ only affects the perturbations. This also explains why increasing $w_{\rm gdm}$ leads to a decrease of $\delta_{\rm c}$: a positive $w_{\rm gdm}$ makes the contribution of the dark matter component less important (as it decreases faster) than that of the cosmological constant at late times and to overcome the additional contribution to the expansion one needs lower overdensities to achieve the collapse.

So far, we have considered the influence of $w_{\rm gdm}$ and $c^2_{\rm s, gdm}$ separately, but to span the full parameter space of the model, we need to consider their combined effect and we do so by solving the full equation of motion (\ref{equ:newt_sc}). We verified that for realistic values of the two parameters, the resulting $\delta_{\rm c}$ is in agreement with $\Lambda$CDM. For the highest values considered for both $w_{\rm gdm}$ and $c^2_{\rm s, gdm}$, the relative difference is still below the percent level.

Our analysis led us to the conclusion that the values allowed for $w_{\rm gdm}$ and $c^2_{\rm s, gdm}$ from previous works on the evolution of linear perturbations (see Table~\ref{table_linear_constraints}) have a negligible impact on $\delta_{\rm c}$ and $\Delta_{\rm V}$. Nevertheless, these two quantities are not directly observable and therefore it is important to study how the mass function is influenced.

\begin{figure*}[!ht]
\centering
\includegraphics[width=0.49\hsize]{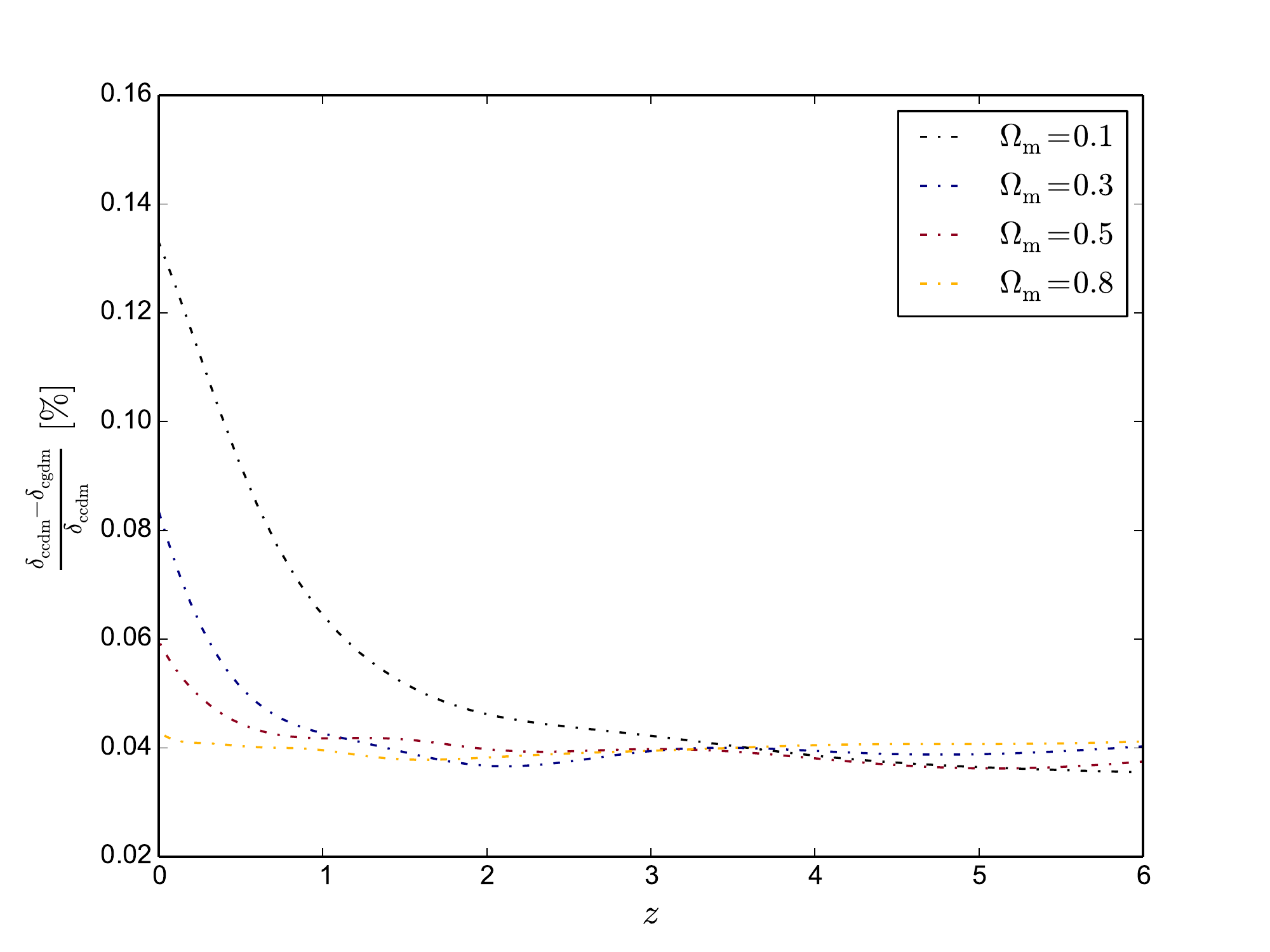}
\includegraphics[width=0.49\hsize]{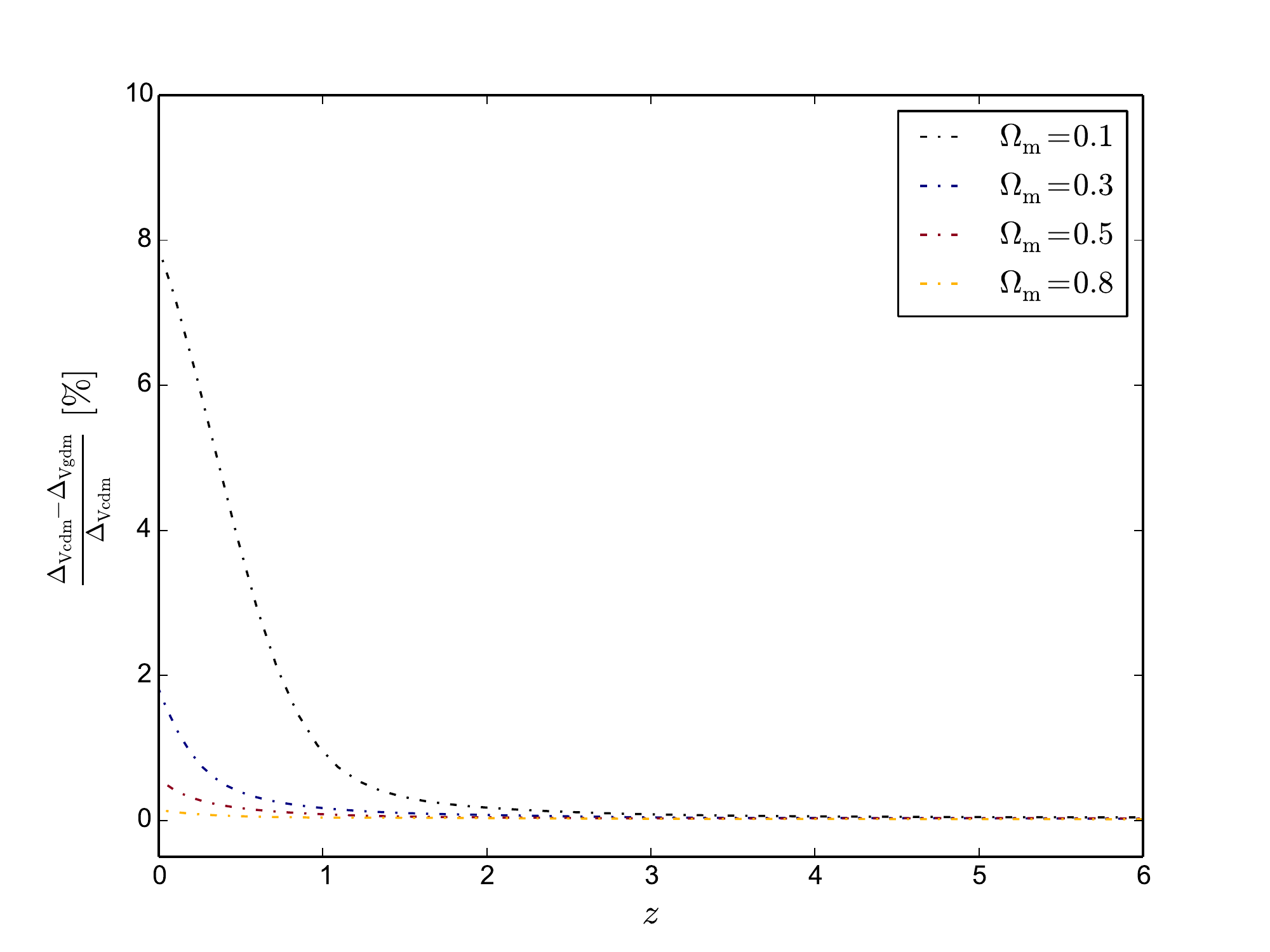}
\caption{Relative differences between the $\Lambda$CDM and $\Lambda$GDM linear critical density parameter $\delta_{\rm c}$ (left) and virial overdensity $\Delta_{\rm V}$ (right) as a function of the collapse redshift $z$ for different values of the matter density parameter $\Omega_{\rm m}$. We assume $w_{\rm gdm}=5\times10^{-4}$ and $c_{\rm s, gdm}^2 = 5\times10^{-7}$ as reference values for the parameters of the GDM model.}
\label{fig:wgcsom}
\end{figure*}

Before we study the impact on the mass function, though, we want to take a step further and investigate the combined action of varying the background matter density parameter $\Omega_{\rm m}$ while fixing the two parameters of the GDM model to $w_{\rm gdm}=5\times10^{-4}$ and $c^2_{\rm s, gdm}=5\times10^{-7}$, as there might be additional degeneracies among the three parameters at the perturbative level. The reason behind this is that $\Omega_{\rm m}$ has a very strong impact on the growth of cosmic structures \cite{Bahcall1997,Blanchard1998} and its determination might be affected when studying a more general dark matter model, as is the case for this work.

In the following, therefore, we vary $\Omega_{\rm m}$ between 0.1 and 0.8 and compare the results for a GDM with the standard CDM model having the same matter density parameter. We present our results in Fig.~\ref{fig:wgcsom}. Stronger effects take place for low matter density parameters for both $\delta_{\rm c}$ and $\Delta_{\rm V}$ as in this case very overdense initial perturbations are required to overcome the accelerated cosmic expansion and collapse. We note that in this case, the effect is of the order of 0.1\% for $\delta_{\rm c}$ and up to 8\% for $\Delta_{\rm V}$. Differences become much smaller at high redshifts, as the EdS is a good approximation of the true cosmology. For a model with $\Omega_{\rm m}\approx 0.3$ (roughly the current accepted value), at $z=0$, we find a 0.08\% difference for $\delta_{\rm c}$ and 2\% for $\Delta_{\rm V}$. Deviations for $\Delta_{\rm V}$ are obviously stronger as this quantity is derived by solving the nonlinear equation of motion for the matter overdensity $\delta$.

\begin{figure}[!ht]
\centering
\includegraphics[width=0.9\hsize]{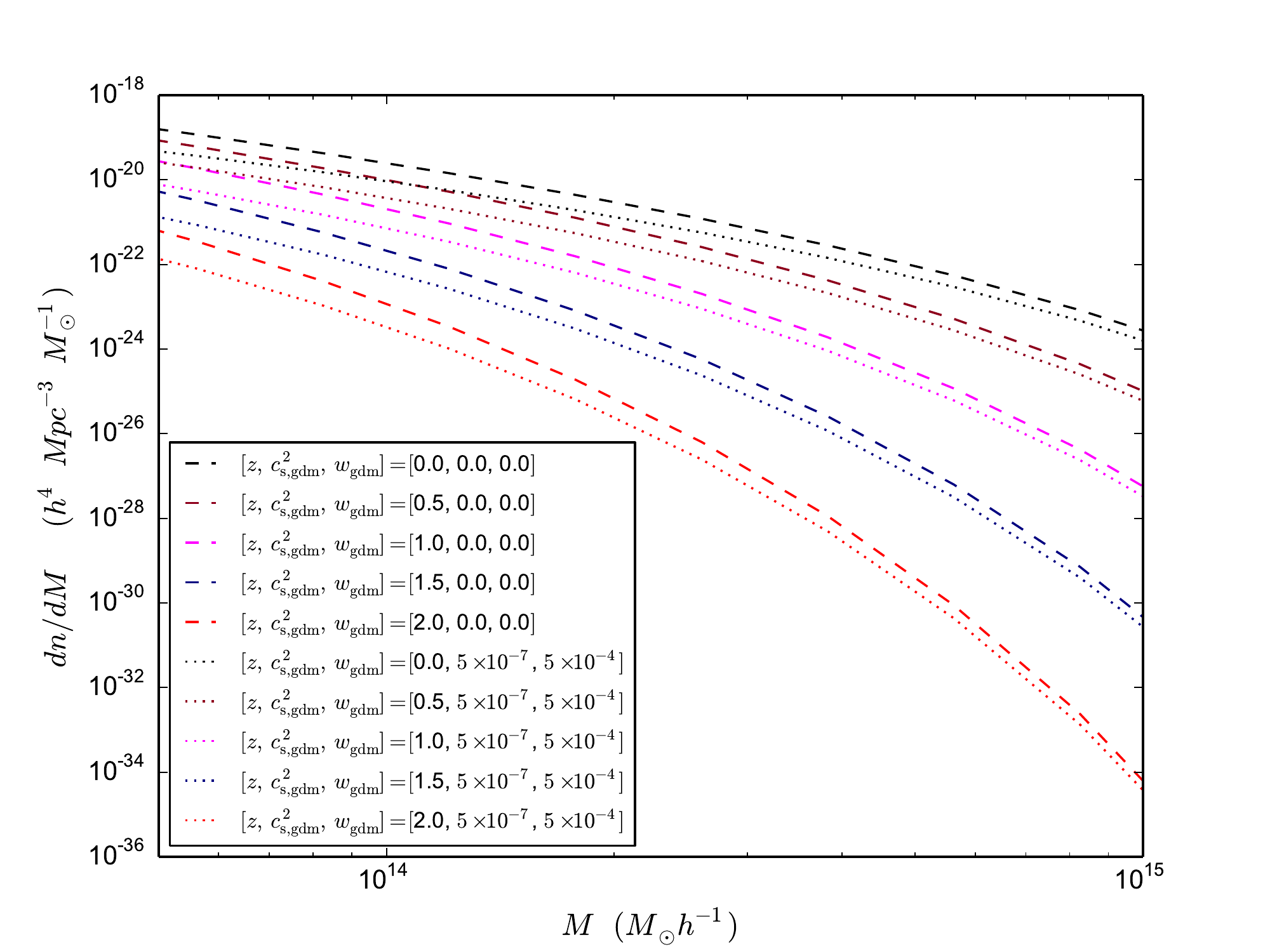}\\
\includegraphics[width=0.9\hsize]{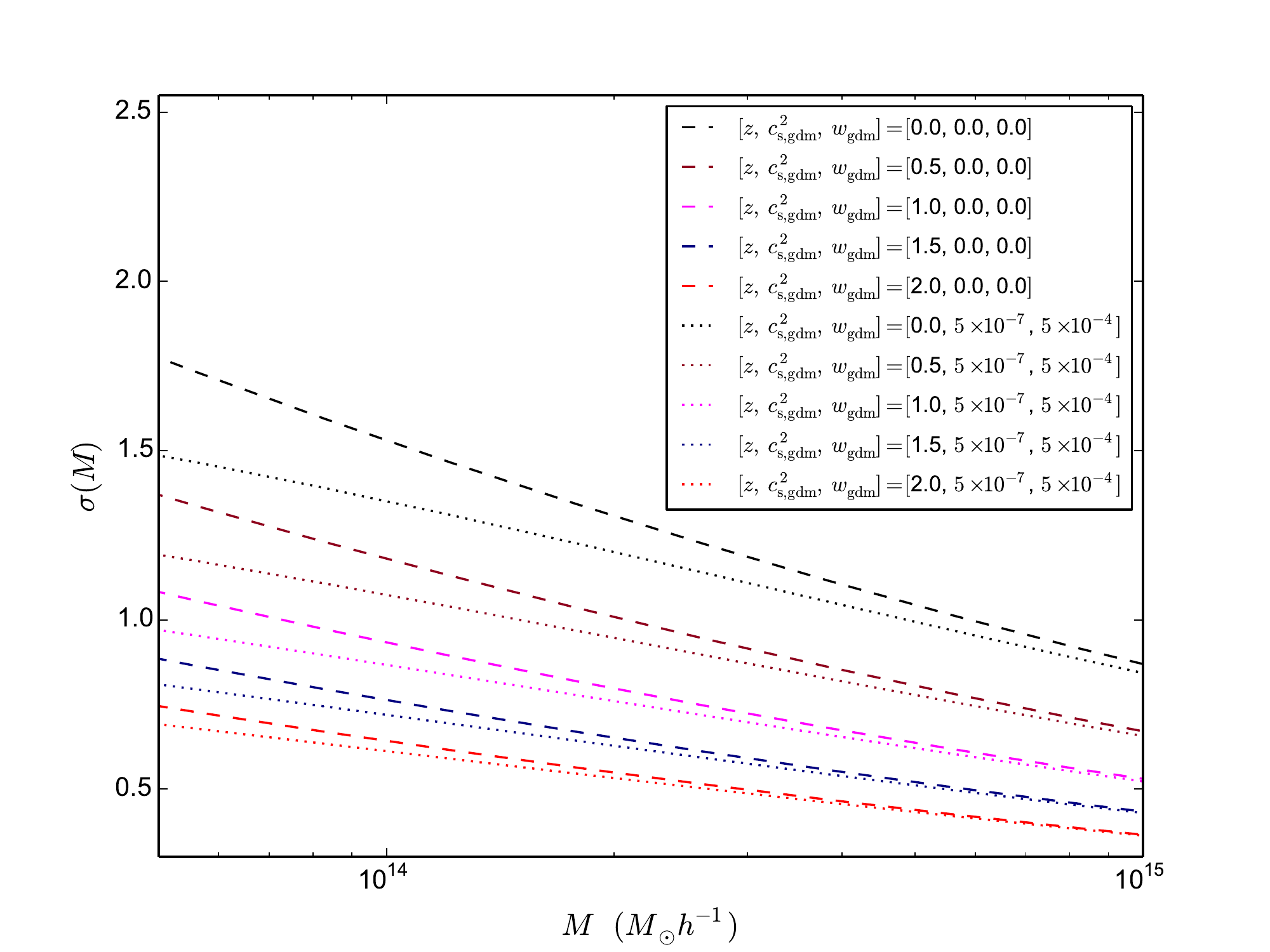}\\
\includegraphics[width=0.9\hsize]{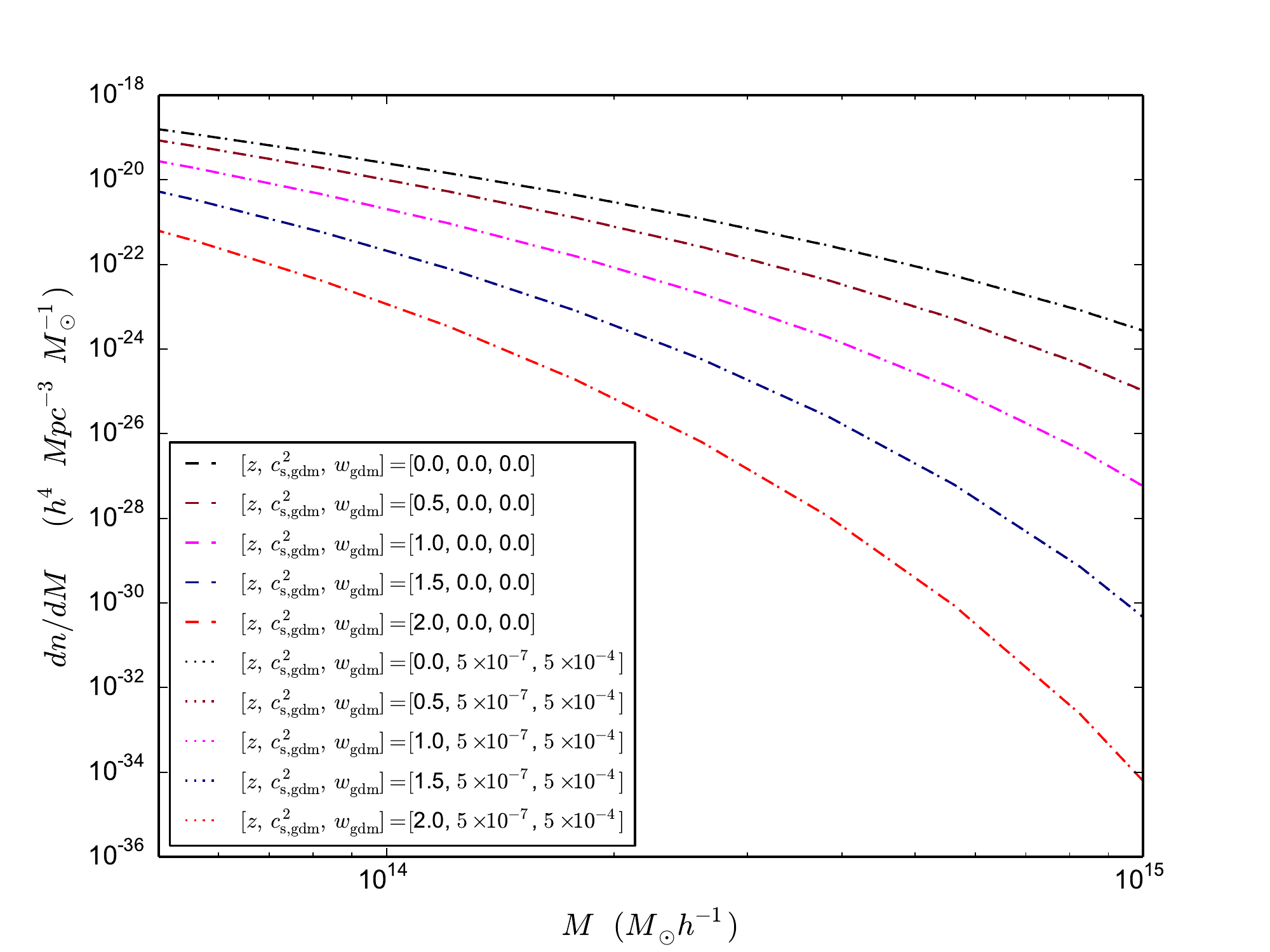}
\caption{In the top panel we show the differential mass function as a function of mass for different redshifts using the appropriate linear matter power spectrum for each model ($\Lambda$CDM and $\Lambda$GDM). In the middle panel we present the evolution of the square root of the mass variance as a function of mass for different redshifts and for both models. In the bottom panel we show the differential mass function as in the top panel but using the $\Lambda$CDM matter power spectrum for both $\Lambda$CDM and $\Lambda$GDM, while the rest of GDM model-dependant quantities are kept. In all panels we assume $w_{\rm gdm}=5\times 10^{-4}$ and $c_{\rm s, gdm}^2=5\times 10^{-7}$ for the $\Lambda$GDM model.}
\label{fig:dndm}
\end{figure}

\subsection{The mass function}
The previous discussion shows that for realistic values of $w_{\rm gdm}$ and $c_{\rm s, gdm}^2$ allowed by previous studies \citep{Thomas2016,Thomas2019,Tutusaus2018}, the parameter to take more into consideration is the equation of state $w_{\rm gdm}$ and that the quantity being mostly affected is the virial overdensity $\Delta_{\rm V}$, with differences up to a few percent for accepted values of the matter density parameter $\Omega_{\rm m}$. Although these numbers are small, we remind the reader that their combined effect enters exponentially into the evaluation of the halo mass function, therefore even small differences can be amplified and lead to appreciable differences, therefore making it a very sensible probe for cosmology when its high-mass end is investigated (i.e., massive galaxy clusters).

The halo mass function is defined as \citep{Blanchard1992}
\begin{equation}\label{equ:nm0}
 \frac{dn(M)}{dM}=-\frac{\bar{\rho}}{M}\frac{\rmd \nu}{\rmd M} \mathcal{F}(\nu)\,,
\end{equation}
where $\bar{\rho}$ is the mean matter density today, $\mathcal{F}(\nu)$ the multiplicity function and $\nu = \delta_{\rm c}/\sigma(M)$ with $\sigma(M)$ the variance within a sphere of radius $R$ and mass $M=4\pi/3\,\bar{\rho}R^3$ for a cosmology described by a linear matter power spectrum $P(k)$. The mass variance is defined as $\sigma^2(M,z)=\frac{1}{2\pi^2}\int_0^\infty k^2 P(k,z) W^2(kR) {\rm d}k$, where $W(kR)$ is an appropriate window function representing the Fourier transform of the top-hat function in real space.

In order to compute the linear matter power spectrum for GDM, we have followed \cite{Thomas2016,Tutusaus2018} in modifying the Einstein-Boltzmann solver CLASS \cite{Lesgourgues2011,Blas2011}.
Thus, at a linear level of perturbations and in the synchronous gauge, the conservation of the energy momentum tensor yields \citep{Ma1995}
\begin{subequations}
\begin{align}
 & \dot{\delta} + (1+w_{\rm gdm})\left( \theta + \frac{\dot{h}}{2} \right) + 
  3H\left(\frac{\delta P}{\delta\rho} -w_{\rm gdm} \right)\delta = 0 \,,\\
 & \dot{\theta} + H(1-3w_{\rm gdm})\theta + \frac{\dot{w}_{\rm gdm}}{1+w_{\rm gdm}} \theta -
 \frac{\delta P/\delta\rho}{1+w_{\rm gdm}} k^2 \delta + \nonumber\\
 & \qquad k^2\sigma = 0 \,.
\end{align}
\end{subequations}
The system is closed by supplying the relations associating the GDM equation of state parameter $w_{\rm gdm}$, the pressure perturbation $\delta P$ and scalar anisotropic stress $\sigma$ to the density fluctuation $\delta$, the divergence of its velocity $\theta$ and the synchronous metric perturbations $h$ and $\eta$,
\begin{align}
 & \delta P = c_{\rm s,gdm}^2 \delta\rho - \dot{\rho} (c_{\rm s,gdm}^2-c_{\rm a,gdm}^2)\theta/k^2\,,\\
 & \dot{\sigma}+3H\frac{c_{\rm a,gdm}^2}{w_{\rm gdm}}\sigma = \frac{4}{3}\frac{c_{\rm vis,gdm}^2}{1+w_{\rm gdm}}(2\theta+\dot{h}+6\dot{\eta}) \label{eq:sigmaEvol}\,,
\end{align}
where the adiabatic sound speed is $c_{\rm a,gdm}^2 \equiv (w_{\rm gdm}\bar{\rho})\dot{}/\dot{\bar{\rho}}$
and $c_{\rm vis,gdm}^2$ is a viscosity parameter we set to zero in this work.
The public version of CLASS already includes a parametrization of a dark energy fluid with constant equation of state parameter and constant sound velocity \cite{Lesgourgues2011b}. We used this parametrization as GDM, while we kept a cosmological constant for the dark energy contribution, and a negligible fraction of CDM. Note that the perturbations of this fluid must then be added to the total matter perturbations, which is not the case in the public version, since this fluid is supposed to behave as dark energy.

For the multiplicity function, we adopt the functional form proposed by \cite{Sheth1999,Sheth2001}
\begin{equation}\label{equ:ST-MF}
 \nu \mathcal{F}_{\rm ST}(\nu) = A\sqrt{\frac{2a}{\pi}}\left[1+\left(\frac{1}{a\nu^2}\right)^p\right]\ \nu \ \exp{\left\{\left[-\frac{a\nu^2}{2}\right]\right\}}\,,
\end{equation}
with the parameters $A$, $a$ and $p$ more recently fitted by \cite{Despali2016} using the abundance matching technique in $N$-body simulations and $\nu=\delta_{\rm c}/(D_{+}\sigma_{\rm M})$, with $\sigma_{\rm M}$ the mass variance. While the mass function depends explicitly on $\delta_{\rm c}$, the fitted parameters are a function of the virial overdensity $\Delta_{\rm V}$ and read
\begin{align*}
 a = &\, 0.4332x^2 + 0.2263x + 0.7665\,,\\
 p = &\, -0.1151x^2 + 0.2554x + 0.2488\,,\\
 A = &\, -0.1362 x + 0.3292\,,
\end{align*}
where $x=\log{(\Delta(z)/\Delta_{\rm V}(z))}$ and $\Delta(z)$ is a given overdensity, such as a multiple of the critical density.\footnote{Note that our definition of the virial overdensity refers to the background density rather than the critical one. Therefore, in the evaluation of the mass function we scale it by $\Omega_{\rm m}$, where necessary.}

Thus, the overall effect on the mass function is given by the combination of a few factors: a different background expansion induced by $w_{\rm gdm}$, the evolution of structures given by the linear growth factor $D_{+}(a)$ and $\delta_{\rm c}$, the evolution of $\Delta_{\rm V}$ and the linear matter power spectrum $P(k)$.

Despite the parameterization adopted here has been originally proposed as an improved fit to $\Lambda$CDM simulations \citep{Sheth1999}, it has been soon afterward justified theoretically based on the ellipsoidal collapse \citep{Sheth2001}. It has been also shown that the Sheth and Tormen parametrization has a more general validity than just the standard $\Lambda$CDM model, as demonstrated by \cite{Pace2014} comparing the results of $N$-body simulations for nonminimally coupled models with theoretical estimations of the halo mass function. The agreement was shown to be very good, provided, though, that the model-dependent parameters (i.e., $\delta_{\rm c}$) were used, rather than the common assumption in the literature where the linear overdensity parameter for the EdS model is used instead of the correct one. According to this, we are confident that our choice is justified.

We present the results of our investigation in the top panel of Fig.~\ref{fig:dndm}, where we show the mass function at different redshifts, considering both $\Lambda$CDM and $\Lambda$GDM models assuming $w_{\rm gdm}=5\times10^{-4}$ and $c_{\rm s, gdm}^2=5\times10^{-7}$ for the latter, two values within the constraints obtained from probes of large scale structure formation \citep{Tutusaus2018,Thomas2019}.

We note immediately that despite $\delta_{\rm c}$ is hardly affected and $\Delta_{\rm V}$ only at the percent level by the combined action of the equation-of-state parameter $w_{\rm gdm}$ and the sound speed $c^2_{\rm s, gdm}$ as we previously discussed, the differential mass function (top panel) shows strong signatures due to the additional physics investigated. One of the main reasons is the strong suppression of power in the linear matter power spectrum due to $c^2_{\rm s, gdm}$ especially on small scales, as shown by \cite{Kopp2016} and as we checked but do not show here by changing the value of $\Delta_{\rm V}$. As we will discuss more in detail later, this directly explains why there is, in general, a lower number of structures, especially for small mass objects. Finally, a part of the contribution could also come from $\Delta_{\rm V}$, as we showed that at the nonlinear level this is the quantity more affected.

More quantitatively, there is a decrement of about 75\% and 80\% for objects of $\approx 5\times10^{13}M_{\odot}\,h^{-1}$ for $z=0$ and $z=2$, respectively. At higher masses, differences between the $\Lambda$CDM and $\Lambda$GDM models are comparable and of the order of 40\%. At low masses though, according to expectations, differences steadily increase with redshift.

To see why we obtain the counterintuitive result of stronger effects at low masses, in the middle panel of Fig.~\ref{fig:dndm}, we show the evolution of the square root of the variance $\sigma(M)$ as a function of the perturbation mass $M$ for different redshifts $z$. We consider both the $\Lambda$CDM and the $\Lambda$GDM models with the same set of parameters we used to study the halo mass function. We immediately see that $\sigma_{\Lambda{\rm GDM}} < \sigma_{\Lambda{\rm CDM}}$ at all masses and redshifts, thus explaining the smaller number of halos in the GDM model. In addition, and this is the key to explain the results for the halo mass function, stronger differences occur at low masses and low redshifts, as the $\Lambda$GDM model approaches $\Lambda$CDM at higher redshifts.

To disentangle the effect of the matter power spectrum from that of the virial overdensity $\Delta_{\rm V}$ entering in the definition of the parameters of the mass function, we evaluate the mass function for $\Lambda$CDM and $\Lambda$GDM by assuming the same $\Lambda$CDM linear matter power spectrum for both models, but keeping the other quantities relative to each model. We show this in the bottom panel of Fig.~\ref{fig:dndm}. 

With respect to before, we now see a completely different situation, which is more in line with usual expectations as the major differences occur, as one would expect, for high-mass objects. Nevertheless, differences are very small and probably more likely due to numerical effects rather than genuine physical effects.

These results thus show that the additional physics of the dark matter sector has a strong impact on the observables, not only on the linear evolution of perturbations but also on the nonlinear evolution of the formation of structure through the halo mass function (top panel). However, within the current linear constraints on $c^2_{\rm s, gdm}$ and $w_{\rm gdm}$ (see Table~\ref{table_linear_constraints}), there is no significant modification coming from changes to $\delta_{\rm c}$ and $\Delta_{\rm V}$, and this is confirmed in the bottom panel where we replaced the correct linear matter power spectrum for the GDM with that expected from the $\Lambda$CDM cosmology.

\subsection{Comparison with observations}
The comparison between the theoretical and the observational halo mass function is not an easy task as one has to take into account complications inferring the halo mass. The determination of the mass can be done via x-ray observations: measuring the x-ray temperature function of galaxy clusters and assuming a mass-temperature relation, it is possible to transform it into a mass function.

This is easy to see noticing that the halo number is a conserved quantity and we can write
\begin{equation}
 n(T,z)\mathrm{d}T = n(M(T,z))\mathrm{d}M\,,
\end{equation}
where $T$ denotes the x-ray temperature. We have, therefore, to establish the relation $\mathrm{d}M/\mathrm{d}T$.

We also point out that this approach needs the catalog used to be flux complete, which is indeed the case for high-mass clusters. But this also implies that a proper selection function should be used to take into account this bias.

For a precise determination of the halo mass function from a given sample catalog, we will use the same procedure outlined in \cite{Campanelli2012}. The catalog used contains massive clusters with $M>8\times 10^{14}\,M_{\odot}\,h^{-1}$ within a comoving radius $R=1.5\,h^{-1}\,{\rm Mpc}$ and span a redshift range $0.05\lesssim z\lesssim 0.83$ which consists of four bins, each with an effective fraction of the observed bin volume (see Table~I of \cite{Campanelli2012}). The x-ray temperatures, as reported in Table~II of \cite{Campanelli2012}, to which we refer for a complete list of the objects used, are taken from \cite{Bahcall1998,Bahcall2003,Ikebe2002,Henry2000,Donahue1998}.

The mass-temperature relation is \cite{Hjorth1998a,Hjorth1998b}
\begin{equation}
 M^{\prime} = 1.5\times 10^{14}\,M_{\odot}\,h^{-1}\kappa_{\Delta}\frac{T_X}{{\rm keV}}\frac{1}{1+z}\,,
\end{equation}
where $M^{\prime}$ is the virial mass contained in a comoving radius $R_0^{\prime}=1.5\,h^{-1}\,{\rm Mpc}$, $T_X$ the cluster x-ray temperature and $\kappa_{\Delta}=0.76$.

Putting this together, the observed number of clusters in a redshift bin $i$ is
\begin{equation}
 \mathcal{N}_i^{\prime} = \alpha_i 
 \int_{z_1^{i}}^{z_2^{i}}\mathrm{d}z\, \frac{\mathrm{d}V}{\mathrm{d}z}
 N^{\prime}(M^{\prime}>M_0^{\prime},z)\,,
\end{equation}
with
\begin{equation}
 N^{\prime}(M^{\prime}>M_0^{\prime},z) = 
 \int_{g(M_0^{\prime})}^{\infty}\mathrm{d}M\, n(M,z)\,,
\end{equation}
and the function $g$ relates the observed mass $M^{\prime}$ to the virial mass and returns the fiducial mass adopted in the observations. We refer the reader to Appendix B of \cite{Campanelli2012} for details on how to evaluate this function. The parameter $\alpha_i$ represents the fraction of the volume observed at that redshift bin. In other words, this means that for a given redshift bin, there is a minimum mass below which the object cannot be detected by the particular survey considered. As the sample is at relatively small redshifts, effects on the selection function due to the cosmology are negligible, taking into account that the background expansion for the GDM model is extremely close to that of $\Lambda$CDM. 
Finally note that this cluster sample was used by \cite{Mehrabi2017} to put constraints on the parameters of the nonspherical collapse model in the $\Lambda$CDM framework.

\begin{figure}[!t]
 \centering
 \includegraphics[width=\hsize]{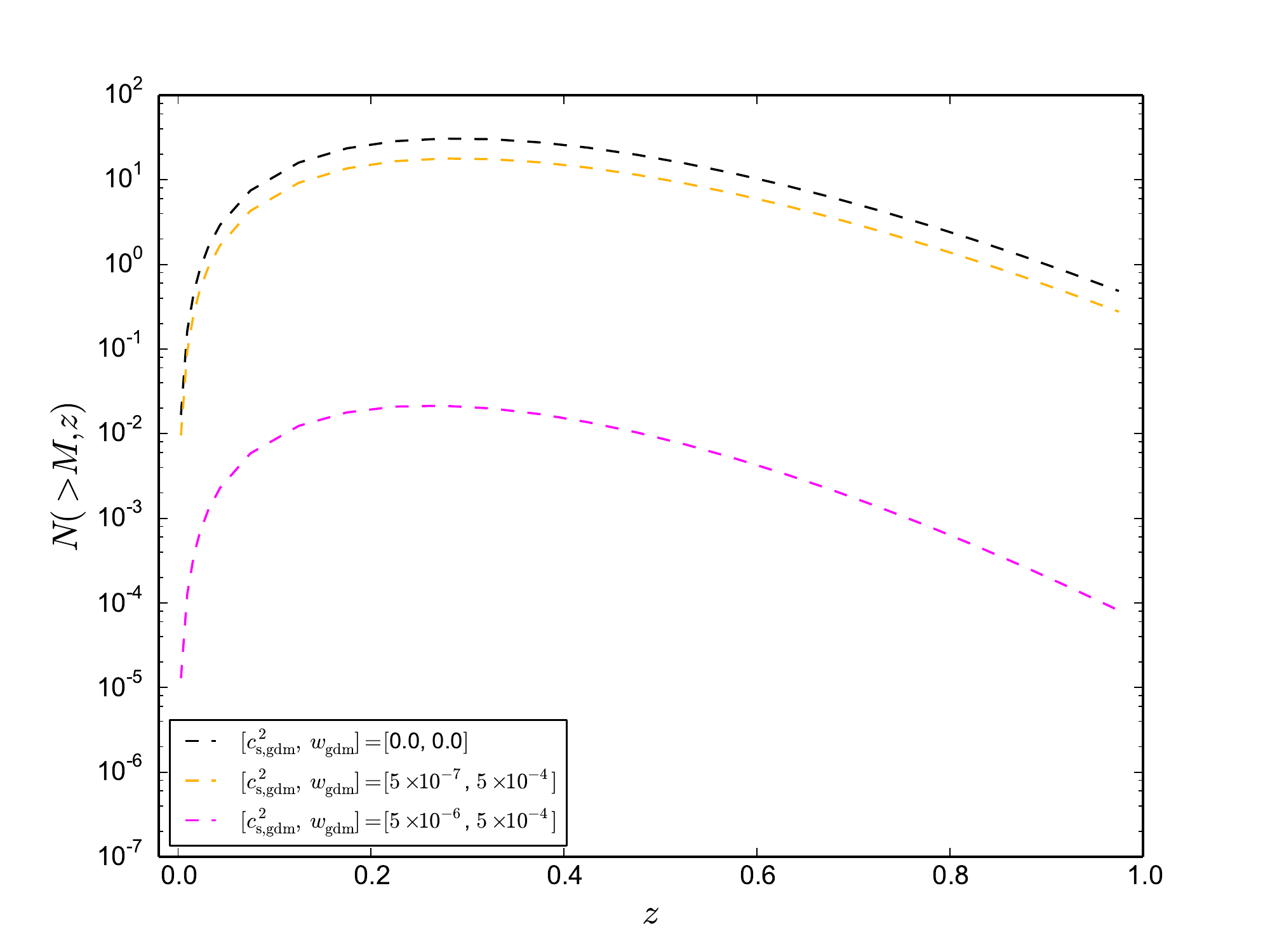}
 \caption{Number of objects above a given mass, as explained in the text, in a given redshift bin. We consider 25 bins between $z=0$ and $z=1$ and three different cosmological models: $\Lambda$CDM (black line), GDM with $w_{\rm gdm}=5\times 10^{-4}$ and $c_{\rm s, gdm}^{2}=5\times 10^{-7}$ (orange line), and GDM with $w_{\rm gdm}=5\times 10^{-4}$ and $c_{\rm s, gdm}^{2}=5\times 10^{-6}$ (purple line).}
 \label{fig:Nobs}
\end{figure}

In Fig.~\ref{fig:Nobs} we present a comparison for the number of objects above a given mass and in a given redshift interval for three different models: the reference $\Lambda$CDM and two GDM models with realistic values for the two parameters $w_{\rm gdm}$ and $c_{\rm s, gdm}^{2}$. For the GDM models, we set $w_{\rm gdm}=5\times 10^{-4}$ and $c_{\rm s, gdm}^2=\{5\times 10^{-7}, 5\times 10^{-6}\}$. We consider 25 redshift bins between the redshift interval $0\leq z \leq 1$ and we show the number of objects with respect to the central value of the bin. It is clear that the sound speed has an important quantitative effect on the halo count, but the shape of the curves is generally the same. The overall effect of increasing the sound speed is a decrease in the number of objects above a given mass, rather independent of the mass. Some stronger effect is seen, as expected, for the high-mass end. From a more quantitative point of view, increasing the sound speed by an order of magnitude leads to about 3 orders of magnitude less objects over the redshift interval investigated. For the GDM model with $c_{\rm s,gdm}^2= 5\times 10^{-6}$ the differences with respect to $\Lambda$CDM are about a factor of 2 and for high redshifts, due to the low number of objects, usually within the uncertainties of the observational probes.

We can now evaluate the $\chi^2$ for the three models discussed in Fig.~\ref{fig:Nobs} using the data and the procedure discussed in \cite{Campanelli2012} constructing a likelihood function $\mathcal{L}$ based on Poisson statistics \citep{Cash1979}
\begin{equation}
 \ln{\mathcal{L}} = \ln{\mathcal{P}(N_i|n_i)} = 
 \sum_{i=1}^{N_{\rm b}}\left[N_i\ln{N_i}-n_i-\ln{(N_i!)}\right]\,,
\end{equation}
where $\mathcal{P}(N_i|n_i)$ is the probability of finding $N_i$ clusters in each of the $n_{\rm b}$ bins given an expected number of $n_i$ in each bin in redshift.

For the $\Lambda$CDM model we find $\chi^2/\text{dof}\approx 6.7/4$ with fixed values of the cosmological parameters ($\Omega_{\rm m}=0.33,\sigma_8=0.81, h=0.675, n_{\rm s}=0.965$), 
while for the GDM models with $c_{\rm s,gdm}^2=5\times 10^{-7}$ and $c_{\rm s,gdm}^2=5\times 10^{-6}$ (and the same cosmological parameters) we find $7.9/4$ and $60.6/4$, respectively. It is immediately clear that the $\Lambda$CDM model is in better agreement with data with respect to these particular $\Lambda$GDM models, and this is a direct consequence of the strong suppression induced by the large sound speeds chosen. It is important to note that this does not disfavor $\Lambda$GDM in any way, since $\Lambda$CDM is a particular case of it. What can be extracted from these results is that values of the GDM parameters currently allowed by measurements in the linear regime (Table
~\ref{table_linear_constraints}), provide a worse $\chi^2$ for GDM. Therefore, small-scales cosmological probes, like cluster counts, will enable to improve our constraints on the GDM parameters and allow us to improve our knowledge on the nature of dark matter.

\subsection{The nonlinear matter power spectrum}
The final part of our investigation deals with the evolution of the matter power spectrum. In general, there are no theoretically motivated procedures which allow us to evaluate the nonlinear evolution of structures. The spherical collapse model is a welcome exception to that and, despite its simplifications, it has proven to be very useful and in very good agreement with results from $N$-body simulations about the evolution of the mass function. This has been shown with detailed studies for $\Lambda$CDM cosmologies comparing the Sheth-Tormen mass function \cite{Sheth1999,Sheth2001} with the Millennium simulation \cite{Springel2005a} or for nonminimally coupled models \cite{Pace2014}.

For the nonlinear matter power spectrum, the situation is rather different. Usually, one can use some fitting function for the $\Lambda$CDM model \cite{Peacock1996} or the halo model \cite{Cooray2002} and adapt them to the particular model considered. This has been done, for example, in \cite{Zhao2014} for $f(R)$ models.

For GDM models, in a recent work \citep{Thomas2019}, the authors developed a formalism in the framework of the halo model to explore how the nonlinear matter power spectrum could evolve. To do so, they used a modified concentration parameter based on the recipe for warm dark matter of \cite{Schneider2012,Marsh2016}. As the halo model also requires the knowledge of the halo mass function, \citep{Thomas2019} related the evolution of $\delta_{\rm c}$ to that of the $\Lambda$CDM model according to the idea that if in GDM models power is removed, then the collapse should be inhibited.

The final result of their machinery is given in their Fig.~2. Nonlinear effects kick in at smaller scales than the $\Lambda$CDM, but the main feature is that also the nonlinear spectrum shows a strong suppression of power. We note that this procedure needs, of course, to be validated with suitably modified $N$-body simulations and it relies, at this stage, on several, albeit plausible, assumptions.

\begin{figure}
 \centering
 \includegraphics[width=\hsize]{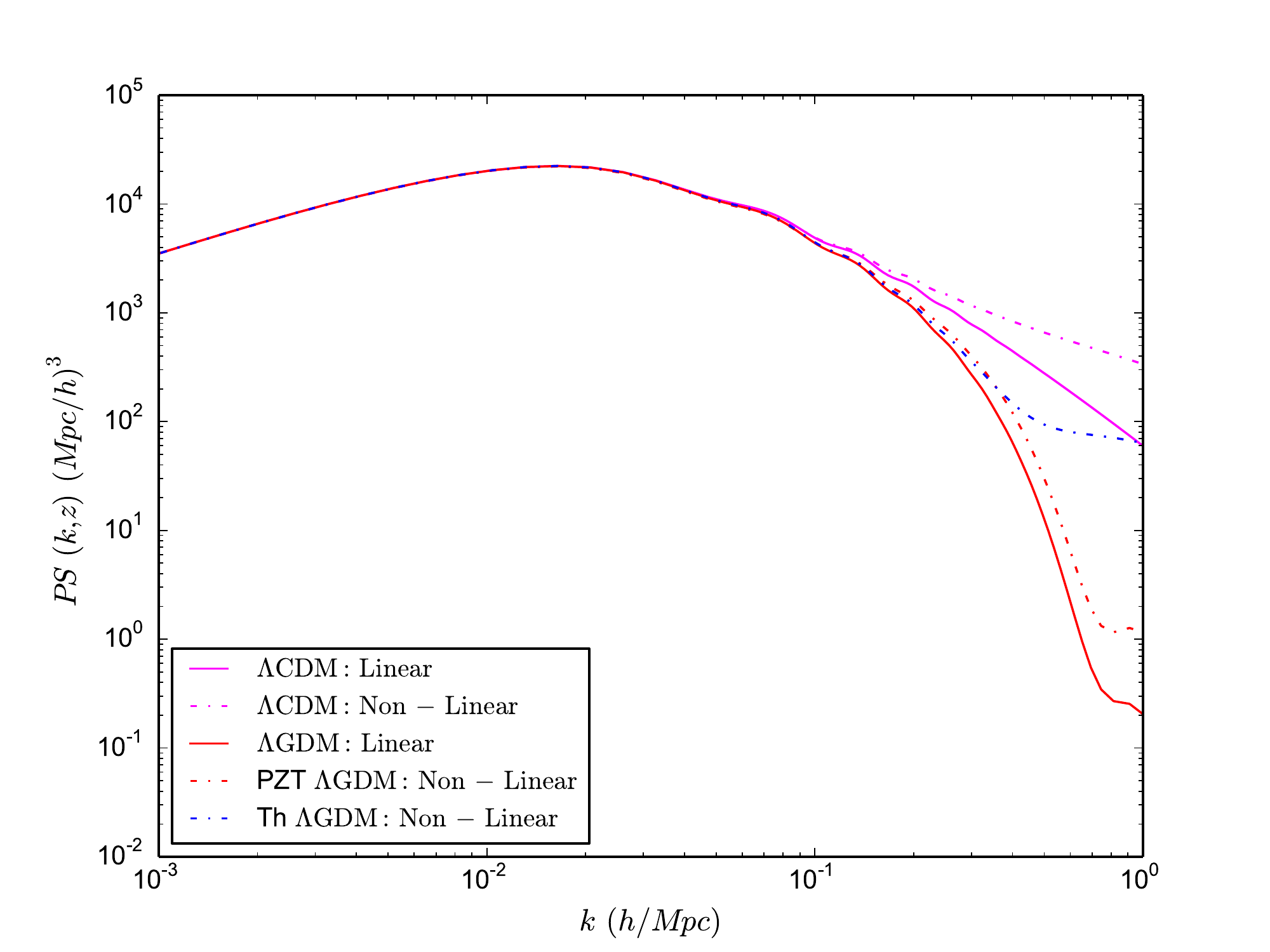}
 \caption{Evolution of the linear (solid lines) and nonlinear (dot-dashed lines) matter power spectra for the $\Lambda$CDM (purple lines) and $\Lambda$GDM (red lines) models at $z=0$. The blue dot-dashed line represents the nonlinear matter power spectrum using the halo-model prescription of \cite{Thomas2019}. For the GDM model, we assumed the following parameters: $w_{\rm gdm}=10^{-4}$ and $c_{\rm s, gdm}^2=10^{-6}$.}
 \label{fig:PS}
\end{figure}

In this work, we, therefore, follow a different approach, which is also used for modified gravity models and does not rely on the knowledge of the nonlinear evolution of perturbations in GDM models, but only on that of a $\Lambda$CDM model. The idea is, in fact, to consider a $\Lambda$CDM model with the same cosmological parameters of the GDM model, determine its nonlinear matter power spectrum using the recipe of \cite{Takahashi2012} and divide it by the linear spectrum. This quantity represents a nonlinear transfer function mapping the linear power spectrum into the nonlinear one. Multiplying the linear GDM matter power spectrum by the nonlinear transfer function, one obtains the corresponding nonlinear evolution of the matter power spectrum. We note that this approach has been used by \cite{SpurioMancini2019} to constrain Horndeski models using cosmic shear, galaxy-galaxy lensing and galaxy clustering with the KIDS \citep{Hildebrandt2017,Kohlinger2017} and GAMA surveys \citep{Driver2009,Driver2011,Liske2015}. As the laws of gravity are not modified, we do not need to take into account further screening mechanisms.

Our results are presented in Fig.~\ref{fig:PS}, where we compare the linear and nonlinear matter power spectrum of a $\Lambda$GDM model with the corresponding $\Lambda$CDM ones at $z=0$. For the GDM model we assumed $w_{\rm gdm}=10^{-4}$ and $c_{\rm s, gdm}^2=10^{-6}$, in agreement with constraints in \cite{Thomas2019}.

With such a low value of the sound speed, the linear spectra of the $\Lambda$CDM and $\Lambda$GDM models agree rather well up to $k\approx 0.02~h{\rm Mpc}^{-1}$, but on smaller scales the new physics kicks in and we easily see a strong suppression in structure, which becomes more and more pronounced the smaller the scale is: for $k\approx 1~h{\rm Mpc}^{-1}$ the two spectra differ by about a factor of 300.

A very similar behaviour is obtained when comparing the nonlinear matter power spectra for the two models. This is easy to understand as the linear GDM spectrum was scaled by the nonlinear transfer function defined above. We remark that this does not have to be the case in general. In addition, it is useful to take into account that our nonlinear mapping is purely phenomenological and, if we exclude the behavior at the linear level, there is no direct dependence on the GDM parameters.

It is therefore important to compare our predictions with the generalized halo model of \cite{Thomas2019}. The two different approaches give very similar results up to $k\approx 0.4~h{\rm Mpc}^{-1}$, but on smaller scales the rescaling method proposed here shows much less power than the halo-model approach discussed in \cite{Thomas2019}. To see whether this is a general feature, we also considered the case with $w_{\rm gdm}=10^{-4}$ and $c_{\rm s, gdm}^2=10^{-5}$. In this case, while our approach still shows an increase of power on small scales, the halo model does not lead to any difference between the linear and nonlinear matter power spectrum, as a high sound speed completely hampers nonlinearities. This probably shows a failure in the regime of validity of the halo model which, though, should not have practical consequences as this value has been already ruled out by linear analysis.

\section{Conclusions}\label{sect:concL}
In this work we discussed how the equation-of-state parameter $w_{\rm gdm}$ and the sound speed $c^2_{\rm s, gdm}$ for generalized dark matter (GDM) affect the properties of the two main quantities of the spherical collapse model, the linear overdensity parameter $\delta_{\rm c}$ and the virial overdensity $\Delta_{\rm V}$.

We compared them with the corresponding quantities derived for the standard $\Lambda$CDM model for different values of $w_{\rm gdm}$ and $c^2_{\rm s, gdm}$. We demonstrated that the parameter mostly affecting their evolution is the background equation of state $w_{\rm gdm}$, while the sound speed $c^2_{\rm s, gdm}$, within the constraints from linear probes, has a negligible contribution to the overall nonlinear evolution.

The effect of the two additional quantities describing the GDM properties is strongly dependent on the matter density parameter $\Omega_{\rm m}$. We saw that the lower the matter density parameter, the stronger are the deviations from a $\Lambda$CDM model, as higher initial overdensities are required to overcome the accelerated expansion of the Universe to allow structures to collapse.

Since the spherical collapse parameters are not direct observables, we used their evolution, together with the linear matter power spectrum $P(k)$ obtained for the GDM model, as building blocks for the halo mass function. We found that major deviations take place on smaller mass objects, rather than at higher masses and this is a direct consequence of the modifications on the linear matter power spectrum, as we verified by using the linear matter power spectrum of the $\Lambda$CDM model for both cosmologies (bottom panel of Fig.~\ref{fig:dndm}). In this case in fact, differences are much smaller and in line with expectations: the decrease is stronger at higher masses, albeit in general at the subpercent level. The overall effect of the GDM dynamics is that of decreasing the number of halos, as additional pressure terms kick in the equations of motion. This is a strong effect, up to 70\% already at $z=0$ and 80\% at $z=2$ and of the order of 40\% for objects of $M\approx10^{15}\,M_{\odot}\,h^{-1}$. Due to the abundance of galactic objects, this effect should be easily seen and therefore put strong constraints on the GDM parameters.

We finally note that, while both important to completely characterize the dynamics of GDM models, the equation-of-state parameter $w_{\rm gdm}$ and the sound speed $c_{\rm s, gdm}^2$ act on two different sets of observables. The first is important mainly at the nonlinear level, while the latter is very important for the evolution of the linear matter power spectrum, which reflects, of course, on the halo mass function. Therefore, while heavily constrained at the linear level already, the two GDM parameters can be further strongly constrained if probes at the nonlinear level are combined.

With the effect on the mass function being detectable with present and most importantly, near-future deep surveys, we remark that a proper analysis of the additional degrees of freedom in a particular cosmological model needs to be taken into full consideration, especially if we do not fix the matter density parameter to the fiducial value. As a practical example for GDM models, a robust measurement of the matter power spectrum and of the mass function, or more generally in the nonlinear regime, can lead to stronger constraints on the GDM parameters.

To this purpose, we implemented a phenomenological approach to study the evolution of the nonlinear matter power spectrum $P(k)$. Knowing the linear and nonlinear matter power spectrum for a $\Lambda$CDM model, we derive a ``nonlinear" transfer function which we multiply with the GDM linear spectrum to infer its nonlinear counterpart. While being approximated, we checked that this approach gives results in good agreement with the more sophisticated approach based on the halo model discussed in \cite{Thomas2019} up to $k\approx 0.4~h{\rm Mpc}^{-1}$ and on small scales a strong lack of power in our approach with respect to \cite{Thomas2019}. For large values of the sound speed, while the approach presented in this work still clearly shows a different evolution between linear and nonlinear spectra, this is not the case anymore for the halo-model approach, as the sound speed completely wipes out any nonlinear evolution. This might, eventually, show a breakdown of the halo model in its current form. We caution, however, that a proper determination of this quantity can be done with accurate $N$-body simulations.

Rather than the differential mass function itself, it is possible to compare theoretical predictions with the number counts one can infer from observed halo catalogs. For that, we compared our predictions with the halo catalog used by \cite{Campanelli2012,Mehrabi2017} to put constraints on the dark energy models and to extensions of the spherical collapse model. We chose the catalog used by \cite{Campanelli2012} as it allows us a study also at higher redshifts, where stronger differences are expected. Studying the $\chi^2$, we showed that $\Lambda$GDM models with GDM parameters allowed by observations in the linear regime provide worse $\chi^2$ values than $\Lambda$CDM. This shows that cluster counts, and especially future larger and deeper surveys, will allow us to put more stringent constraints on the nature of dark matter. 
Note that an increase in our dark matter knowledge is very important, since we showed that a change in the values of GDM parameters could translate into changes of large scale structure abundance. These could then help to explain recent discrepancies between local and deep large scale structure measurements of the growth of structures, which common modifications to $\Lambda$CDM failed to achieve \citep{Ilic2019}.

\section*{Acknowledgements}
F.P. acknowledges financial support from the STFC Grant No. R120562 and thanks Daniel Thomas for useful discussions on the halo-model approach for the GDM models and for providing the nonlinear matter spectrum for a quantitative comparison with our approach. The authors also thank the referees for their useful comments which helped to improve the scientific content of the paper.

\bibliographystyle{apsrev4-1}
\bibliography{GDMSC.bib}

\begin{thebibliography}{63}%
\makeatletter
\providecommand \@ifxundefined [1]{%
 \@ifx{#1\undefined}
}%
\providecommand \@ifnum [1]{%
 \ifnum #1\expandafter \@firstoftwo
 \else \expandafter \@secondoftwo
 \fi
}%
\providecommand \@ifx [1]{%
 \ifx #1\expandafter \@firstoftwo
 \else \expandafter \@secondoftwo
 \fi
}%
\providecommand \natexlab [1]{#1}%
\providecommand \enquote  [1]{``#1''}%
\providecommand \bibnamefont  [1]{#1}%
\providecommand \bibfnamefont [1]{#1}%
\providecommand \citenamefont [1]{#1}%
\providecommand \href@noop [0]{\@secondoftwo}%
\providecommand \href [0]{\begingroup \@sanitize@url \@href}%
\providecommand \@href[1]{\@@startlink{#1}\@@href}%
\providecommand \@@href[1]{\endgroup#1\@@endlink}%
\providecommand \@sanitize@url [0]{\catcode `\\12\catcode `\$12\catcode
  `\&12\catcode `\#12\catcode `\^12\catcode `\_12\catcode `\%12\relax}%
\providecommand \@@startlink[1]{}%
\providecommand \@@endlink[0]{}%
\providecommand \url  [0]{\begingroup\@sanitize@url \@url }%
\providecommand \@url [1]{\endgroup\@href {#1}{\urlprefix }}%
\providecommand \urlprefix  [0]{URL }%
\providecommand \Eprint [0]{\href }%
\providecommand \doibase [0]{http://dx.doi.org/}%
\providecommand \selectlanguage [0]{\@gobble}%
\providecommand \bibinfo  [0]{\@secondoftwo}%
\providecommand \bibfield  [0]{\@secondoftwo}%
\providecommand \translation [1]{[#1]}%
\providecommand \BibitemOpen [0]{}%
\providecommand \bibitemStop [0]{}%
\providecommand \bibitemNoStop [0]{.\EOS\space}%
\providecommand \EOS [0]{\spacefactor3000\relax}%
\providecommand \BibitemShut  [1]{\csname bibitem#1\endcsname}%
\let\auto@bib@innerbib\@empty
\bibitem [{\citenamefont {{Frieman}}\ \emph {et~al.}(2008)\citenamefont
  {{Frieman}}, \citenamefont {{Turner}},\ and\ \citenamefont
  {{Huterer}}}]{Frieman2008}%
  \BibitemOpen
  \bibfield  {author} {\bibinfo {author} {\bibfnamefont {J.~A.}\ \bibnamefont
  {{Frieman}}}, \bibinfo {author} {\bibfnamefont {M.~S.}\ \bibnamefont
  {{Turner}}}, \ and\ \bibinfo {author} {\bibfnamefont {D.}~\bibnamefont
  {{Huterer}}},\ }\href {\doibase 10.1146/annurev.astro.46.060407.145243}
  {\bibfield  {journal} {\bibinfo  {journal} {\araa}\ }\textbf {\bibinfo
  {volume} {46}},\ \bibinfo {pages} {385} (\bibinfo {year} {2008})},\ \Eprint
  {http://arxiv.org/abs/0803.0982} {arXiv:0803.0982} \BibitemShut {NoStop}%
\bibitem [{\citenamefont {{Spergel}}(2015)}]{Spergel2015}%
  \BibitemOpen
  \bibfield  {author} {\bibinfo {author} {\bibfnamefont {D.~N.}\ \bibnamefont
  {{Spergel}}},\ }\href {\doibase 10.1126/science.aaa0980} {\bibfield
  {journal} {\bibinfo  {journal} {Science}\ }\textbf {\bibinfo {volume}
  {347}},\ \bibinfo {pages} {1100} (\bibinfo {year} {2015})}\BibitemShut
  {NoStop}%
\bibitem [{\citenamefont {{Planck Collaboration VI}}(2018)}]{Planck2018_VI}%
  \BibitemOpen
  \bibfield  {author} {\bibinfo {author} {\bibnamefont {{Planck Collaboration
  VI}}},\ }\href@noop {} {\bibfield  {journal} {\bibinfo  {journal} {ArXiv
  e-prints}\ } (\bibinfo {year} {2018})},\ \Eprint
  {http://arxiv.org/abs/1807.06209} {arXiv:1807.06209} \BibitemShut {NoStop}%
\bibitem [{\citenamefont {{DESI Collaboration}}(2016)}]{DESI2016}%
  \BibitemOpen
  \bibfield  {author} {\bibinfo {author} {\bibnamefont {{DESI
  Collaboration}}},\ }\href@noop {} {\bibfield  {journal} {\bibinfo  {journal}
  {arXiv e-prints}\ ,\ \bibinfo {eid} {arXiv:1611.00036}} (\bibinfo {year}
  {2016})},\ \Eprint {http://arxiv.org/abs/1611.00036} {arXiv:1611.00036
  [astro-ph.IM]} \BibitemShut {NoStop}%
\bibitem [{\citenamefont {{Laureijs}}\ \emph {et~al.}(2011)\citenamefont
  {{Laureijs}}, \citenamefont {{Amiaux}}, \citenamefont {{Arduini}},
  \citenamefont {{Augu{\`e}res}}, \citenamefont {{Brinchmann}}, \citenamefont
  {{Cole}}, \citenamefont {{Cropper}}, \citenamefont {{Dabin}}, \citenamefont
  {{Duvet}}, \citenamefont {{Ealet}},\ and\ \citenamefont
  {et~al.}}]{Laureijs2011}%
  \BibitemOpen
  \bibfield  {author} {\bibinfo {author} {\bibfnamefont {R.}~\bibnamefont
  {{Laureijs}}}, \bibinfo {author} {\bibfnamefont {J.}~\bibnamefont
  {{Amiaux}}}, \bibinfo {author} {\bibfnamefont {S.}~\bibnamefont {{Arduini}}},
  \bibinfo {author} {\bibfnamefont {J.~.}\ \bibnamefont {{Augu{\`e}res}}},
  \bibinfo {author} {\bibfnamefont {J.}~\bibnamefont {{Brinchmann}}}, \bibinfo
  {author} {\bibfnamefont {R.}~\bibnamefont {{Cole}}}, \bibinfo {author}
  {\bibfnamefont {M.}~\bibnamefont {{Cropper}}}, \bibinfo {author}
  {\bibfnamefont {C.}~\bibnamefont {{Dabin}}}, \bibinfo {author} {\bibfnamefont
  {L.}~\bibnamefont {{Duvet}}}, \bibinfo {author} {\bibfnamefont
  {A.}~\bibnamefont {{Ealet}}}, \ and\ \bibinfo {author} {\bibnamefont
  {et~al.}},\ }\href@noop {} {\bibfield  {journal} {\bibinfo  {journal} {ArXiv
  e-prints, 1110.3193}\ } (\bibinfo {year} {2011})},\ \Eprint
  {http://arxiv.org/abs/1110.3193} {arXiv:1110.3193 [astro-ph.CO]} \BibitemShut
  {NoStop}%
\bibitem [{\citenamefont {{Marshall}}\ \emph {et~al.}(2017)\citenamefont
  {{Marshall}}, \citenamefont {{Clarkson}}, \citenamefont {{Shemmer}},
  \citenamefont {{Biswas}}, \citenamefont {{de Val-Borro}}, \citenamefont
  {{Rho}}, \citenamefont {{Jones}}, \citenamefont {{Anguita}}, \citenamefont
  {{Ridgway}},\ and\ \citenamefont {{et~al.}}}]{Marshall2017}%
  \BibitemOpen
  \bibfield  {author} {\bibinfo {author} {\bibfnamefont {P.}~\bibnamefont
  {{Marshall}}}, \bibinfo {author} {\bibfnamefont {W.}~\bibnamefont
  {{Clarkson}}}, \bibinfo {author} {\bibfnamefont {O.}~\bibnamefont
  {{Shemmer}}}, \bibinfo {author} {\bibfnamefont {R.}~\bibnamefont {{Biswas}}},
  \bibinfo {author} {\bibfnamefont {M.}~\bibnamefont {{de Val-Borro}}},
  \bibinfo {author} {\bibfnamefont {J.}~\bibnamefont {{Rho}}}, \bibinfo
  {author} {\bibfnamefont {L.}~\bibnamefont {{Jones}}}, \bibinfo {author}
  {\bibfnamefont {T.}~\bibnamefont {{Anguita}}}, \bibinfo {author}
  {\bibfnamefont {S.}~\bibnamefont {{Ridgway}}}, \ and\ \bibinfo {author}
  {\bibnamefont {{et~al.}}},\ }\href {\doibase 10.5281/zenodo.842713} {\enquote
  {\bibinfo {title} {{Lsst Science Collaborations Observing Strategy White
  Paper: ``Science-Driven Optimization Of The Lsst Observing Strategy''}},}\ }
  (\bibinfo {year} {2017})\BibitemShut {NoStop}%
\bibitem [{\citenamefont {{Dore}}\ \emph {et~al.}(2019)\citenamefont {{Dore}},
  \citenamefont {{Hirata}}, \citenamefont {{Wang}}, \citenamefont {{Weinberg}},
  \citenamefont {{Eifler}}, \citenamefont {{Foley}}, \citenamefont
  {{Heinrich}}, \citenamefont {{Krause}}, \citenamefont {{Perlmutter}},\ and\
  \citenamefont {{et~al.}}}]{Dore2019}%
  \BibitemOpen
  \bibfield  {author} {\bibinfo {author} {\bibfnamefont {O.}~\bibnamefont
  {{Dore}}}, \bibinfo {author} {\bibfnamefont {C.}~\bibnamefont {{Hirata}}},
  \bibinfo {author} {\bibfnamefont {Y.}~\bibnamefont {{Wang}}}, \bibinfo
  {author} {\bibfnamefont {D.}~\bibnamefont {{Weinberg}}}, \bibinfo {author}
  {\bibfnamefont {T.}~\bibnamefont {{Eifler}}}, \bibinfo {author}
  {\bibfnamefont {R.~J.}\ \bibnamefont {{Foley}}}, \bibinfo {author}
  {\bibfnamefont {C.~H.}\ \bibnamefont {{Heinrich}}}, \bibinfo {author}
  {\bibfnamefont {E.}~\bibnamefont {{Krause}}}, \bibinfo {author}
  {\bibfnamefont {S.}~\bibnamefont {{Perlmutter}}}, \ and\ \bibinfo {author}
  {\bibnamefont {{et~al.}}},\ }\href@noop {} {\bibfield  {journal} {\bibinfo
  {journal} {\baas}\ }\textbf {\bibinfo {volume} {51}},\ \bibinfo {eid} {341}
  (\bibinfo {year} {2019})},\ \Eprint {http://arxiv.org/abs/1904.01174}
  {arXiv:1904.01174 [astro-ph.CO]} \BibitemShut {NoStop}%
\bibitem [{\citenamefont {{Square Kilometre Array Cosmology Science Working
  Group}}\ and\ \citenamefont {{et~al.}}(2018)}]{SKA2018}%
  \BibitemOpen
  \bibfield  {author} {\bibinfo {author} {\bibnamefont {{Square Kilometre Array
  Cosmology Science Working Group}}}\ and\ \bibinfo {author} {\bibnamefont
  {{et~al.}}},\ }\href@noop {} {\bibfield  {journal} {\bibinfo  {journal}
  {arXiv e-prints}\ ,\ \bibinfo {eid} {arXiv:1811.02743}} (\bibinfo {year}
  {2018})},\ \Eprint {http://arxiv.org/abs/1811.02743} {arXiv:1811.02743
  [astro-ph.CO]} \BibitemShut {NoStop}%
\bibitem [{\citenamefont {{Cooray}}\ and\ \citenamefont
  {{Sheth}}(2002)}]{Cooray2002}%
  \BibitemOpen
  \bibfield  {author} {\bibinfo {author} {\bibfnamefont {A.}~\bibnamefont
  {{Cooray}}}\ and\ \bibinfo {author} {\bibfnamefont {R.}~\bibnamefont
  {{Sheth}}},\ }\href {\doibase 10.1016/S0370-1573(02)00276-4} {\bibfield
  {journal} {\bibinfo  {journal} {\physrep}\ }\textbf {\bibinfo {volume}
  {372}},\ \bibinfo {pages} {1} (\bibinfo {year} {2002})},\ \Eprint
  {http://arxiv.org/abs/arXiv:astro-ph/0206508} {arXiv:astro-ph/0206508}
  \BibitemShut {NoStop}%
\bibitem [{\citenamefont {{Gunn}}\ and\ \citenamefont
  {{Gott}}(1972)}]{Gunn1972}%
  \BibitemOpen
  \bibfield  {author} {\bibinfo {author} {\bibfnamefont {J.~E.}\ \bibnamefont
  {{Gunn}}}\ and\ \bibinfo {author} {\bibfnamefont {J.~R.}\ \bibnamefont
  {{Gott}}, \bibfnamefont {III}},\ }\href {\doibase 10.1086/151605} {\bibfield
  {journal} {\bibinfo  {journal} {\apj}\ }\textbf {\bibinfo {volume} {176}},\
  \bibinfo {pages} {1} (\bibinfo {year} {1972})}\BibitemShut {NoStop}%
\bibitem [{\citenamefont {{Wang}}\ and\ \citenamefont
  {{Steinhardt}}(1998)}]{Wang1998}%
  \BibitemOpen
  \bibfield  {author} {\bibinfo {author} {\bibfnamefont {L.}~\bibnamefont
  {{Wang}}}\ and\ \bibinfo {author} {\bibfnamefont {P.~J.}\ \bibnamefont
  {{Steinhardt}}},\ }\href {\doibase 10.1086/306436} {\bibfield  {journal}
  {\bibinfo  {journal} {\apj}\ }\textbf {\bibinfo {volume} {508}},\ \bibinfo
  {pages} {483} (\bibinfo {year} {1998})},\ \Eprint
  {http://arxiv.org/abs/arXiv:astro-ph/9804015} {arXiv:astro-ph/9804015}
  \BibitemShut {NoStop}%
\bibitem [{\citenamefont {{Horellou}}\ and\ \citenamefont
  {{Berge}}(2005)}]{Horellou2005}%
  \BibitemOpen
  \bibfield  {author} {\bibinfo {author} {\bibfnamefont {C.}~\bibnamefont
  {{Horellou}}}\ and\ \bibinfo {author} {\bibfnamefont {J.}~\bibnamefont
  {{Berge}}},\ }\href {\doibase 10.1111/j.1365-2966.2005.09140.x} {\bibfield
  {journal} {\bibinfo  {journal} {\mnras}\ }\textbf {\bibinfo {volume} {360}},\
  \bibinfo {pages} {1393} (\bibinfo {year} {2005})},\ \Eprint
  {http://arxiv.org/abs/arXiv:astro-ph/0504465} {arXiv:astro-ph/0504465}
  \BibitemShut {NoStop}%
\bibitem [{\citenamefont {{Pace}}\ \emph {et~al.}(2010)\citenamefont {{Pace}},
  \citenamefont {{Waizmann}},\ and\ \citenamefont {{Bartelmann}}}]{Pace2010}%
  \BibitemOpen
  \bibfield  {author} {\bibinfo {author} {\bibfnamefont {F.}~\bibnamefont
  {{Pace}}}, \bibinfo {author} {\bibfnamefont {J.-C.}\ \bibnamefont
  {{Waizmann}}}, \ and\ \bibinfo {author} {\bibfnamefont {M.}~\bibnamefont
  {{Bartelmann}}},\ }\href {\doibase 10.1111/j.1365-2966.2010.16841.x}
  {\bibfield  {journal} {\bibinfo  {journal} {\mnras}\ }\textbf {\bibinfo
  {volume} {406}},\ \bibinfo {pages} {1865} (\bibinfo {year} {2010})},\ \Eprint
  {http://arxiv.org/abs/1005.0233} {arXiv:1005.0233 [astro-ph.CO]} \BibitemShut
  {NoStop}%
\bibitem [{\citenamefont {{Pace}}\ \emph {et~al.}(2012)\citenamefont {{Pace}},
  \citenamefont {{Fedeli}}, \citenamefont {{Moscardini}},\ and\ \citenamefont
  {{Bartelmann}}}]{Pace2012}%
  \BibitemOpen
  \bibfield  {author} {\bibinfo {author} {\bibfnamefont {F.}~\bibnamefont
  {{Pace}}}, \bibinfo {author} {\bibfnamefont {C.}~\bibnamefont {{Fedeli}}},
  \bibinfo {author} {\bibfnamefont {L.}~\bibnamefont {{Moscardini}}}, \ and\
  \bibinfo {author} {\bibfnamefont {M.}~\bibnamefont {{Bartelmann}}},\ }\href
  {\doibase 10.1111/j.1365-2966.2012.20692.x} {\bibfield  {journal} {\bibinfo
  {journal} {\mnras}\ }\textbf {\bibinfo {volume} {422}},\ \bibinfo {pages}
  {1186} (\bibinfo {year} {2012})},\ \Eprint {http://arxiv.org/abs/1111.1556}
  {arXiv:1111.1556} \BibitemShut {NoStop}%
\bibitem [{\citenamefont {{Pace}}\ \emph {et~al.}(2014)\citenamefont {{Pace}},
  \citenamefont {{Moscardini}}, \citenamefont {{Crittenden}}, \citenamefont
  {{Bartelmann}},\ and\ \citenamefont {{Pettorino}}}]{Pace2014}%
  \BibitemOpen
  \bibfield  {author} {\bibinfo {author} {\bibfnamefont {F.}~\bibnamefont
  {{Pace}}}, \bibinfo {author} {\bibfnamefont {L.}~\bibnamefont
  {{Moscardini}}}, \bibinfo {author} {\bibfnamefont {R.}~\bibnamefont
  {{Crittenden}}}, \bibinfo {author} {\bibfnamefont {M.}~\bibnamefont
  {{Bartelmann}}}, \ and\ \bibinfo {author} {\bibfnamefont {V.}~\bibnamefont
  {{Pettorino}}},\ }\href {\doibase 10.1093/mnras/stt1907} {\bibfield
  {journal} {\bibinfo  {journal} {\mnras}\ }\textbf {\bibinfo {volume} {437}},\
  \bibinfo {pages} {547} (\bibinfo {year} {2014})},\ \Eprint
  {http://arxiv.org/abs/1307.7026} {arXiv:1307.7026} \BibitemShut {NoStop}%
\bibitem [{\citenamefont {{Pace}}\ \emph {et~al.}(2019)\citenamefont {{Pace}},
  \citenamefont {{Schimd}}, \citenamefont {{Mota}},\ and\ \citenamefont {{Del
  Popolo}}}]{Pace2019b}%
  \BibitemOpen
  \bibfield  {author} {\bibinfo {author} {\bibfnamefont {F.}~\bibnamefont
  {{Pace}}}, \bibinfo {author} {\bibfnamefont {C.}~\bibnamefont {{Schimd}}},
  \bibinfo {author} {\bibfnamefont {D.~F.}\ \bibnamefont {{Mota}}}, \ and\
  \bibinfo {author} {\bibfnamefont {A.}~\bibnamefont {{Del Popolo}}},\ }\href
  {\doibase 10.1088/1475-7516/2019/09/060} {\bibfield  {journal} {\bibinfo
  {journal} {\jcap}\ }\textbf {\bibinfo {volume} {2019}},\ \bibinfo {eid} {060}
  (\bibinfo {year} {2019})},\ \Eprint {http://arxiv.org/abs/1811.12105}
  {arXiv:1811.12105} \BibitemShut {NoStop}%
\bibitem [{\citenamefont {{Batista}}\ and\ \citenamefont
  {{Pace}}(2013)}]{Batista2013}%
  \BibitemOpen
  \bibfield  {author} {\bibinfo {author} {\bibfnamefont {R.~C.}\ \bibnamefont
  {{Batista}}}\ and\ \bibinfo {author} {\bibfnamefont {F.}~\bibnamefont
  {{Pace}}},\ }\href {\doibase 10.1088/1475-7516/2013/06/044} {\bibfield
  {journal} {\bibinfo  {journal} {\jcap}\ }\textbf {\bibinfo {volume} {6}},\
  \bibinfo {eid} {044} (\bibinfo {year} {2013})},\ \Eprint
  {http://arxiv.org/abs/1303.0414} {arXiv:1303.0414 [astro-ph.CO]} \BibitemShut
  {NoStop}%
\bibitem [{\citenamefont {{Nazari-Pooya}}\ \emph {et~al.}(2016)\citenamefont
  {{Nazari-Pooya}}, \citenamefont {{Malekjani}}, \citenamefont {{Pace}},\ and\
  \citenamefont {{Jassur}}}]{NazariPooya2016}%
  \BibitemOpen
  \bibfield  {author} {\bibinfo {author} {\bibfnamefont {N.}~\bibnamefont
  {{Nazari-Pooya}}}, \bibinfo {author} {\bibfnamefont {M.}~\bibnamefont
  {{Malekjani}}}, \bibinfo {author} {\bibfnamefont {F.}~\bibnamefont {{Pace}}},
  \ and\ \bibinfo {author} {\bibfnamefont {D.~M.-Z.}\ \bibnamefont
  {{Jassur}}},\ }\href {\doibase 10.1093/mnras/stw582} {\bibfield  {journal}
  {\bibinfo  {journal} {\mnras}\ }\textbf {\bibinfo {volume} {458}},\ \bibinfo
  {pages} {3795} (\bibinfo {year} {2016})},\ \Eprint
  {http://arxiv.org/abs/1601.04593} {arXiv:1601.04593} \BibitemShut {NoStop}%
\bibitem [{\citenamefont {{Pace}}\ \emph {et~al.}(2017)\citenamefont {{Pace}},
  \citenamefont {{Meyer}},\ and\ \citenamefont {{Bartelmann}}}]{Pace2017a}%
  \BibitemOpen
  \bibfield  {author} {\bibinfo {author} {\bibfnamefont {F.}~\bibnamefont
  {{Pace}}}, \bibinfo {author} {\bibfnamefont {S.}~\bibnamefont {{Meyer}}}, \
  and\ \bibinfo {author} {\bibfnamefont {M.}~\bibnamefont {{Bartelmann}}},\
  }\href {\doibase 10.1088/1475-7516/2017/10/040} {\bibfield  {journal}
  {\bibinfo  {journal} {\jcap}\ }\textbf {\bibinfo {volume} {10}},\ \bibinfo
  {eid} {040} (\bibinfo {year} {2017})},\ \Eprint
  {http://arxiv.org/abs/1708.02477} {arXiv:1708.02477} \BibitemShut {NoStop}%
\bibitem [{\citenamefont {{Hu}}(1998)}]{Hu1998}%
  \BibitemOpen
  \bibfield  {author} {\bibinfo {author} {\bibfnamefont {W.}~\bibnamefont
  {{Hu}}},\ }\href {\doibase 10.1086/306274} {\bibfield  {journal} {\bibinfo
  {journal} {\apj}\ }\textbf {\bibinfo {volume} {506}},\ \bibinfo {pages} {485}
  (\bibinfo {year} {1998})},\ \Eprint {http://arxiv.org/abs/astro-ph/9801234}
  {astro-ph/9801234} \BibitemShut {NoStop}%
\bibitem [{\citenamefont {{Kopp}}\ \emph {et~al.}(2016)\citenamefont {{Kopp}},
  \citenamefont {{Skordis}},\ and\ \citenamefont {{Thomas}}}]{Kopp2016}%
  \BibitemOpen
  \bibfield  {author} {\bibinfo {author} {\bibfnamefont {M.}~\bibnamefont
  {{Kopp}}}, \bibinfo {author} {\bibfnamefont {C.}~\bibnamefont {{Skordis}}}, \
  and\ \bibinfo {author} {\bibfnamefont {D.~B.}\ \bibnamefont {{Thomas}}},\
  }\href {\doibase 10.1103/PhysRevD.94.043512} {\bibfield  {journal} {\bibinfo
  {journal} {\prd}\ }\textbf {\bibinfo {volume} {94}},\ \bibinfo {eid} {043512}
  (\bibinfo {year} {2016})},\ \Eprint {http://arxiv.org/abs/1605.00649}
  {arXiv:1605.00649 [astro-ph.CO]} \BibitemShut {NoStop}%
\bibitem [{\citenamefont {{Kunz}}\ \emph {et~al.}(2016)\citenamefont {{Kunz}},
  \citenamefont {{Nesseris}},\ and\ \citenamefont {{Sawicki}}}]{Kunz2016}%
  \BibitemOpen
  \bibfield  {author} {\bibinfo {author} {\bibfnamefont {M.}~\bibnamefont
  {{Kunz}}}, \bibinfo {author} {\bibfnamefont {S.}~\bibnamefont {{Nesseris}}},
  \ and\ \bibinfo {author} {\bibfnamefont {I.}~\bibnamefont {{Sawicki}}},\
  }\href {\doibase 10.1103/PhysRevD.94.023510} {\bibfield  {journal} {\bibinfo
  {journal} {\prd}\ }\textbf {\bibinfo {volume} {94}},\ \bibinfo {eid} {023510}
  (\bibinfo {year} {2016})},\ \Eprint {http://arxiv.org/abs/1604.05701}
  {arXiv:1604.05701 [astro-ph.CO]} \BibitemShut {NoStop}%
\bibitem [{\citenamefont {{Thomas}}\ \emph {et~al.}(2016)\citenamefont
  {{Thomas}}, \citenamefont {{Kopp}},\ and\ \citenamefont
  {{Skordis}}}]{Thomas2016}%
  \BibitemOpen
  \bibfield  {author} {\bibinfo {author} {\bibfnamefont {D.~B.}\ \bibnamefont
  {{Thomas}}}, \bibinfo {author} {\bibfnamefont {M.}~\bibnamefont {{Kopp}}}, \
  and\ \bibinfo {author} {\bibfnamefont {C.}~\bibnamefont {{Skordis}}},\ }\href
  {\doibase 10.3847/0004-637X/830/2/155} {\bibfield  {journal} {\bibinfo
  {journal} {\apj}\ }\textbf {\bibinfo {volume} {830}},\ \bibinfo {eid} {155}
  (\bibinfo {year} {2016})},\ \Eprint {http://arxiv.org/abs/1601.05097}
  {arXiv:1601.05097 [astro-ph.CO]} \BibitemShut {NoStop}%
\bibitem [{\citenamefont {{Kopp}}\ \emph {et~al.}(2018)\citenamefont {{Kopp}},
  \citenamefont {{Skordis}}, \citenamefont {{Thomas}},\ and\ \citenamefont
  {{Ili{\'c}}}}]{Kopp2018}%
  \BibitemOpen
  \bibfield  {author} {\bibinfo {author} {\bibfnamefont {M.}~\bibnamefont
  {{Kopp}}}, \bibinfo {author} {\bibfnamefont {C.}~\bibnamefont {{Skordis}}},
  \bibinfo {author} {\bibfnamefont {D.~B.}\ \bibnamefont {{Thomas}}}, \ and\
  \bibinfo {author} {\bibfnamefont {S.}~\bibnamefont {{Ili{\'c}}}},\ }\href
  {\doibase 10.1103/PhysRevLett.120.221102} {\bibfield  {journal} {\bibinfo
  {journal} {\prl}\ }\textbf {\bibinfo {volume} {120}},\ \bibinfo {eid}
  {221102} (\bibinfo {year} {2018})},\ \Eprint
  {http://arxiv.org/abs/1802.09541} {arXiv:1802.09541 [astro-ph.CO]}
  \BibitemShut {NoStop}%
\bibitem [{\citenamefont {{Tutusaus}}\ \emph {et~al.}(2018)\citenamefont
  {{Tutusaus}}, \citenamefont {{Lamine}},\ and\ \citenamefont
  {{Blanchard}}}]{Tutusaus2018}%
  \BibitemOpen
  \bibfield  {author} {\bibinfo {author} {\bibfnamefont {I.}~\bibnamefont
  {{Tutusaus}}}, \bibinfo {author} {\bibfnamefont {B.}~\bibnamefont
  {{Lamine}}}, \ and\ \bibinfo {author} {\bibfnamefont {A.}~\bibnamefont
  {{Blanchard}}},\ }\href@noop {} {\bibfield  {journal} {\bibinfo  {journal}
  {arXiv e-prints}\ ,\ \bibinfo {eid} {arXiv:1805.06202}} (\bibinfo {year}
  {2018})},\ \Eprint {http://arxiv.org/abs/1805.06202} {arXiv:1805.06202
  [astro-ph.CO]} \BibitemShut {NoStop}%
\bibitem [{\citenamefont {{Thomas}}\ \emph {et~al.}(2019)\citenamefont
  {{Thomas}}, \citenamefont {{Kopp}},\ and\ \citenamefont
  {{Markovi{\v{c}}}}}]{Thomas2019}%
  \BibitemOpen
  \bibfield  {author} {\bibinfo {author} {\bibfnamefont {D.~B.}\ \bibnamefont
  {{Thomas}}}, \bibinfo {author} {\bibfnamefont {M.}~\bibnamefont {{Kopp}}}, \
  and\ \bibinfo {author} {\bibfnamefont {K.}~\bibnamefont {{Markovi{\v{c}}}}},\
  }\href {\doibase 10.1093/mnras/stz2559} {\bibfield  {journal} {\bibinfo
  {journal} {\mnras}\ }\textbf {\bibinfo {volume} {490}},\ \bibinfo {pages}
  {813} (\bibinfo {year} {2019})},\ \Eprint {http://arxiv.org/abs/1905.02739}
  {arXiv:1905.02739 [astro-ph.CO]} \BibitemShut {NoStop}%
\bibitem [{\citenamefont {{Abramo}}\ \emph {et~al.}(2007)\citenamefont
  {{Abramo}}, \citenamefont {{Batista}}, \citenamefont {{Liberato}},\ and\
  \citenamefont {{Rosenfeld}}}]{Abramo2007}%
  \BibitemOpen
  \bibfield  {author} {\bibinfo {author} {\bibfnamefont {L.~R.}\ \bibnamefont
  {{Abramo}}}, \bibinfo {author} {\bibfnamefont {R.~C.}\ \bibnamefont
  {{Batista}}}, \bibinfo {author} {\bibfnamefont {L.}~\bibnamefont
  {{Liberato}}}, \ and\ \bibinfo {author} {\bibfnamefont {R.}~\bibnamefont
  {{Rosenfeld}}},\ }\href {\doibase 10.1088/1475-7516/2007/11/012} {\bibfield
  {journal} {\bibinfo  {journal} {Journal of Cosmology and Astro-Particle
  Physics}\ }\textbf {\bibinfo {volume} {11}},\ \bibinfo {pages} {12} (\bibinfo
  {year} {2007})},\ \Eprint {http://arxiv.org/abs/0707.2882} {arXiv:0707.2882}
  \BibitemShut {NoStop}%
\bibitem [{\citenamefont {{Lahav}}\ \emph {et~al.}(1991)\citenamefont
  {{Lahav}}, \citenamefont {{Lilje}}, \citenamefont {{Primack}},\ and\
  \citenamefont {{Rees}}}]{Lahav1991}%
  \BibitemOpen
  \bibfield  {author} {\bibinfo {author} {\bibfnamefont {O.}~\bibnamefont
  {{Lahav}}}, \bibinfo {author} {\bibfnamefont {P.~B.}\ \bibnamefont
  {{Lilje}}}, \bibinfo {author} {\bibfnamefont {J.~R.}\ \bibnamefont
  {{Primack}}}, \ and\ \bibinfo {author} {\bibfnamefont {M.~J.}\ \bibnamefont
  {{Rees}}},\ }\href@noop {} {\bibfield  {journal} {\bibinfo  {journal}
  {\mnras}\ }\textbf {\bibinfo {volume} {251}},\ \bibinfo {pages} {128}
  (\bibinfo {year} {1991})}\BibitemShut {NoStop}%
\bibitem [{\citenamefont {{Landau}}\ and\ \citenamefont
  {{Lifshitz}}(1969)}]{Mechanics1969}%
  \BibitemOpen
  \bibfield  {author} {\bibinfo {author} {\bibfnamefont {L.~D.}\ \bibnamefont
  {{Landau}}}\ and\ \bibinfo {author} {\bibfnamefont {E.~M.}\ \bibnamefont
  {{Lifshitz}}},\ }\href@noop {} {\emph {\bibinfo {title} {{Mechanics}}}}\
  (\bibinfo {year} {Pergamon Press, 1969})\BibitemShut {NoStop}%
\bibitem [{\citenamefont {{Iliev}}\ and\ \citenamefont
  {{Shapiro}}(2001)}]{Iliev2001}%
  \BibitemOpen
  \bibfield  {author} {\bibinfo {author} {\bibfnamefont {I.~T.}\ \bibnamefont
  {{Iliev}}}\ and\ \bibinfo {author} {\bibfnamefont {P.~R.}\ \bibnamefont
  {{Shapiro}}},\ }\href {\doibase 10.1046/j.1365-8711.2001.04422.x} {\bibfield
  {journal} {\bibinfo  {journal} {\mnras}\ }\textbf {\bibinfo {volume} {325}},\
  \bibinfo {pages} {468} (\bibinfo {year} {2001})},\ \Eprint
  {http://arxiv.org/abs/astro-ph/0101067} {astro-ph/0101067} \BibitemShut
  {NoStop}%
\bibitem [{\citenamefont {{Bahcall}}\ and\ \citenamefont
  {{Fan}}(1997)}]{Bahcall1997}%
  \BibitemOpen
  \bibfield  {author} {\bibinfo {author} {\bibfnamefont {N.~A.}\ \bibnamefont
  {{Bahcall}}}\ and\ \bibinfo {author} {\bibfnamefont {X.}~\bibnamefont
  {{Fan}}},\ }in\ \href@noop {} {\emph {\bibinfo {booktitle} {American
  Astronomical Society Meeting Abstracts}}},\ \bibinfo {series} {American
  Astronomical Society Meeting Abstracts}, Vol.\ \bibinfo {volume} {191}\
  (\bibinfo {year} {1997})\ p.\ \bibinfo {pages} {95.02}\BibitemShut {NoStop}%
\bibitem [{\citenamefont {{Blanchard}}\ and\ \citenamefont
  {{Bartlett}}(1998)}]{Blanchard1998}%
  \BibitemOpen
  \bibfield  {author} {\bibinfo {author} {\bibfnamefont {A.}~\bibnamefont
  {{Blanchard}}}\ and\ \bibinfo {author} {\bibfnamefont {J.~G.}\ \bibnamefont
  {{Bartlett}}},\ }\href@noop {} {\bibfield  {journal} {\bibinfo  {journal}
  {\aap}\ }\textbf {\bibinfo {volume} {332}},\ \bibinfo {pages} {L49} (\bibinfo
  {year} {1998})},\ \Eprint {http://arxiv.org/abs/astro-ph/9712078}
  {arXiv:astro-ph/9712078 [astro-ph]} \BibitemShut {NoStop}%
\bibitem [{\citenamefont {{Blanchard}}\ \emph {et~al.}(1992)\citenamefont
  {{Blanchard}}, \citenamefont {{Valls-Gabaud}},\ and\ \citenamefont
  {{Mamon}}}]{Blanchard1992}%
  \BibitemOpen
  \bibfield  {author} {\bibinfo {author} {\bibfnamefont {A.}~\bibnamefont
  {{Blanchard}}}, \bibinfo {author} {\bibfnamefont {D.}~\bibnamefont
  {{Valls-Gabaud}}}, \ and\ \bibinfo {author} {\bibfnamefont {G.~A.}\
  \bibnamefont {{Mamon}}},\ }\href@noop {} {\bibfield  {journal} {\bibinfo
  {journal} {\aap}\ }\textbf {\bibinfo {volume} {264}},\ \bibinfo {pages} {365}
  (\bibinfo {year} {1992})}\BibitemShut {NoStop}%
\bibitem [{\citenamefont {{Lesgourgues}}(2011)}]{Lesgourgues2011}%
  \BibitemOpen
  \bibfield  {author} {\bibinfo {author} {\bibfnamefont {J.}~\bibnamefont
  {{Lesgourgues}}},\ }\href@noop {} {\bibfield  {journal} {\bibinfo  {journal}
  {ArXiv e-prints}\ } (\bibinfo {year} {2011})},\ \Eprint
  {http://arxiv.org/abs/1104.2932} {arXiv:1104.2932} \BibitemShut {NoStop}%
\bibitem [{\citenamefont {{Blas}}\ \emph {et~al.}(2011)\citenamefont {{Blas}},
  \citenamefont {{Lesgourgues}},\ and\ \citenamefont {{Tram}}}]{Blas2011}%
  \BibitemOpen
  \bibfield  {author} {\bibinfo {author} {\bibfnamefont {D.}~\bibnamefont
  {{Blas}}}, \bibinfo {author} {\bibfnamefont {J.}~\bibnamefont
  {{Lesgourgues}}}, \ and\ \bibinfo {author} {\bibfnamefont {T.}~\bibnamefont
  {{Tram}}},\ }\href {\doibase 10.1088/1475-7516/2011/07/034} {\bibfield
  {journal} {\bibinfo  {journal} {\jcap}\ }\textbf {\bibinfo {volume} {2011}},\
  \bibinfo {eid} {034} (\bibinfo {year} {2011})},\ \Eprint
  {http://arxiv.org/abs/1104.2933} {arXiv:1104.2933 [astro-ph.CO]} \BibitemShut
  {NoStop}%
\bibitem [{\citenamefont {{Ma}}\ and\ \citenamefont
  {{Bertschinger}}(1995)}]{Ma1995}%
  \BibitemOpen
  \bibfield  {author} {\bibinfo {author} {\bibfnamefont {C.-P.}\ \bibnamefont
  {{Ma}}}\ and\ \bibinfo {author} {\bibfnamefont {E.}~\bibnamefont
  {{Bertschinger}}},\ }\href {\doibase 10.1086/176550} {\bibfield  {journal}
  {\bibinfo  {journal} {\apj}\ }\textbf {\bibinfo {volume} {455}},\ \bibinfo
  {pages} {7} (\bibinfo {year} {1995})},\ \Eprint
  {http://arxiv.org/abs/astro-ph/9506072} {arXiv:astro-ph/9506072 [astro-ph]}
  \BibitemShut {NoStop}%
\bibitem [{\citenamefont {{Lesgourgues}}\ and\ \citenamefont
  {{Tram}}(2011)}]{Lesgourgues2011b}%
  \BibitemOpen
  \bibfield  {author} {\bibinfo {author} {\bibfnamefont {J.}~\bibnamefont
  {{Lesgourgues}}}\ and\ \bibinfo {author} {\bibfnamefont {T.}~\bibnamefont
  {{Tram}}},\ }\href {\doibase 10.1088/1475-7516/2011/09/032} {\bibfield
  {journal} {\bibinfo  {journal} {\jcap}\ }\textbf {\bibinfo {volume} {2011}},\
  \bibinfo {eid} {032} (\bibinfo {year} {2011})},\ \Eprint
  {http://arxiv.org/abs/1104.2935} {arXiv:1104.2935 [astro-ph.CO]} \BibitemShut
  {NoStop}%
\bibitem [{\citenamefont {{Sheth}}\ and\ \citenamefont
  {{Tormen}}(1999)}]{Sheth1999}%
  \BibitemOpen
  \bibfield  {author} {\bibinfo {author} {\bibfnamefont {R.~K.}\ \bibnamefont
  {{Sheth}}}\ and\ \bibinfo {author} {\bibfnamefont {G.}~\bibnamefont
  {{Tormen}}},\ }\href {\doibase 10.1046/j.1365-8711.1999.02692.x} {\bibfield
  {journal} {\bibinfo  {journal} {\mnras}\ }\textbf {\bibinfo {volume} {308}},\
  \bibinfo {pages} {119} (\bibinfo {year} {1999})},\ \Eprint
  {http://arxiv.org/abs/arXiv:astro-ph/9901122} {arXiv:astro-ph/9901122}
  \BibitemShut {NoStop}%
\bibitem [{\citenamefont {{Sheth}}\ \emph {et~al.}(2001)\citenamefont
  {{Sheth}}, \citenamefont {{Mo}},\ and\ \citenamefont {{Tormen}}}]{Sheth2001}%
  \BibitemOpen
  \bibfield  {author} {\bibinfo {author} {\bibfnamefont {R.~K.}\ \bibnamefont
  {{Sheth}}}, \bibinfo {author} {\bibfnamefont {H.~J.}\ \bibnamefont {{Mo}}}, \
  and\ \bibinfo {author} {\bibfnamefont {G.}~\bibnamefont {{Tormen}}},\ }\href
  {\doibase 10.1046/j.1365-8711.2001.04006.x} {\bibfield  {journal} {\bibinfo
  {journal} {\mnras}\ }\textbf {\bibinfo {volume} {323}},\ \bibinfo {pages} {1}
  (\bibinfo {year} {2001})},\ \Eprint
  {http://arxiv.org/abs/arXiv:astro-ph/9907024} {arXiv:astro-ph/9907024}
  \BibitemShut {NoStop}%
\bibitem [{\citenamefont {{Despali}}\ \emph {et~al.}(2016)\citenamefont
  {{Despali}}, \citenamefont {{Giocoli}}, \citenamefont {{Angulo}},
  \citenamefont {{Tormen}}, \citenamefont {{Sheth}}, \citenamefont {{Baso}},\
  and\ \citenamefont {{Moscardini}}}]{Despali2016}%
  \BibitemOpen
  \bibfield  {author} {\bibinfo {author} {\bibfnamefont {G.}~\bibnamefont
  {{Despali}}}, \bibinfo {author} {\bibfnamefont {C.}~\bibnamefont
  {{Giocoli}}}, \bibinfo {author} {\bibfnamefont {R.~E.}\ \bibnamefont
  {{Angulo}}}, \bibinfo {author} {\bibfnamefont {G.}~\bibnamefont {{Tormen}}},
  \bibinfo {author} {\bibfnamefont {R.~K.}\ \bibnamefont {{Sheth}}}, \bibinfo
  {author} {\bibfnamefont {G.}~\bibnamefont {{Baso}}}, \ and\ \bibinfo {author}
  {\bibfnamefont {L.}~\bibnamefont {{Moscardini}}},\ }\href {\doibase
  10.1093/mnras/stv2842} {\bibfield  {journal} {\bibinfo  {journal} {\mnras}\
  }\textbf {\bibinfo {volume} {456}},\ \bibinfo {pages} {2486} (\bibinfo {year}
  {2016})},\ \Eprint {http://arxiv.org/abs/1507.05627} {arXiv:1507.05627}
  \BibitemShut {NoStop}%
\bibitem [{\citenamefont {{Campanelli}}\ \emph {et~al.}(2012)\citenamefont
  {{Campanelli}}, \citenamefont {{Fogli}}, \citenamefont {{Kahniashvili}},
  \citenamefont {{Marrone}},\ and\ \citenamefont {{Ratra}}}]{Campanelli2012}%
  \BibitemOpen
  \bibfield  {author} {\bibinfo {author} {\bibfnamefont {L.}~\bibnamefont
  {{Campanelli}}}, \bibinfo {author} {\bibfnamefont {G.~L.}\ \bibnamefont
  {{Fogli}}}, \bibinfo {author} {\bibfnamefont {T.}~\bibnamefont
  {{Kahniashvili}}}, \bibinfo {author} {\bibfnamefont {A.}~\bibnamefont
  {{Marrone}}}, \ and\ \bibinfo {author} {\bibfnamefont {B.}~\bibnamefont
  {{Ratra}}},\ }\href {\doibase 10.1140/epjc/s10052-012-2218-4} {\bibfield
  {journal} {\bibinfo  {journal} {European Physical Journal C}\ }\textbf
  {\bibinfo {volume} {72}},\ \bibinfo {eid} {2218} (\bibinfo {year} {2012})},\
  \Eprint {http://arxiv.org/abs/1110.2310} {arXiv:1110.2310 [astro-ph.CO]}
  \BibitemShut {NoStop}%
\bibitem [{\citenamefont {{Bahcall}}\ and\ \citenamefont
  {{Fan}}(1998)}]{Bahcall1998}%
  \BibitemOpen
  \bibfield  {author} {\bibinfo {author} {\bibfnamefont {N.~A.}\ \bibnamefont
  {{Bahcall}}}\ and\ \bibinfo {author} {\bibfnamefont {X.}~\bibnamefont
  {{Fan}}},\ }\href {\doibase 10.1086/306088} {\bibfield  {journal} {\bibinfo
  {journal} {\apj}\ }\textbf {\bibinfo {volume} {504}},\ \bibinfo {pages} {1}
  (\bibinfo {year} {1998})},\ \Eprint {http://arxiv.org/abs/astro-ph/9803277}
  {arXiv:astro-ph/9803277 [astro-ph]} \BibitemShut {NoStop}%
\bibitem [{\citenamefont {{Bahcall}}\ and\ \citenamefont
  {{Bode}}(2003)}]{Bahcall2003}%
  \BibitemOpen
  \bibfield  {author} {\bibinfo {author} {\bibfnamefont {N.~A.}\ \bibnamefont
  {{Bahcall}}}\ and\ \bibinfo {author} {\bibfnamefont {P.}~\bibnamefont
  {{Bode}}},\ }\href {\doibase 10.1086/375503} {\bibfield  {journal} {\bibinfo
  {journal} {\apjl}\ }\textbf {\bibinfo {volume} {588}},\ \bibinfo {pages} {L1}
  (\bibinfo {year} {2003})},\ \Eprint {http://arxiv.org/abs/astro-ph/0212363}
  {arXiv:astro-ph/0212363 [astro-ph]} \BibitemShut {NoStop}%
\bibitem [{\citenamefont {{Ikebe}}\ \emph {et~al.}(2002)\citenamefont
  {{Ikebe}}, \citenamefont {{Reiprich}}, \citenamefont {{B{\"o}hringer}},
  \citenamefont {{Tanaka}},\ and\ \citenamefont {{Kitayama}}}]{Ikebe2002}%
  \BibitemOpen
  \bibfield  {author} {\bibinfo {author} {\bibfnamefont {Y.}~\bibnamefont
  {{Ikebe}}}, \bibinfo {author} {\bibfnamefont {T.~H.}\ \bibnamefont
  {{Reiprich}}}, \bibinfo {author} {\bibfnamefont {H.}~\bibnamefont
  {{B{\"o}hringer}}}, \bibinfo {author} {\bibfnamefont {Y.}~\bibnamefont
  {{Tanaka}}}, \ and\ \bibinfo {author} {\bibfnamefont {T.}~\bibnamefont
  {{Kitayama}}},\ }\href {\doibase 10.1051/0004-6361:20011769} {\bibfield
  {journal} {\bibinfo  {journal} {\aap}\ }\textbf {\bibinfo {volume} {383}},\
  \bibinfo {pages} {773} (\bibinfo {year} {2002})},\ \Eprint
  {http://arxiv.org/abs/astro-ph/0112315} {arXiv:astro-ph/0112315 [astro-ph]}
  \BibitemShut {NoStop}%
\bibitem [{\citenamefont {{Henry}}(2000)}]{Henry2000}%
  \BibitemOpen
  \bibfield  {author} {\bibinfo {author} {\bibfnamefont {J.~P.}\ \bibnamefont
  {{Henry}}},\ }\href {\doibase 10.1086/308783} {\bibfield  {journal} {\bibinfo
   {journal} {\apj}\ }\textbf {\bibinfo {volume} {534}},\ \bibinfo {pages}
  {565} (\bibinfo {year} {2000})},\ \Eprint
  {http://arxiv.org/abs/astro-ph/0002365} {arXiv:astro-ph/0002365 [astro-ph]}
  \BibitemShut {NoStop}%
\bibitem [{\citenamefont {{Donahue}}\ \emph {et~al.}(1998)\citenamefont
  {{Donahue}}, \citenamefont {{Voit}}, \citenamefont {{Gioia}}, \citenamefont
  {{Luppino}}, \citenamefont {{Hughes}},\ and\ \citenamefont
  {{Stocke}}}]{Donahue1998}%
  \BibitemOpen
  \bibfield  {author} {\bibinfo {author} {\bibfnamefont {M.}~\bibnamefont
  {{Donahue}}}, \bibinfo {author} {\bibfnamefont {G.~M.}\ \bibnamefont
  {{Voit}}}, \bibinfo {author} {\bibfnamefont {I.}~\bibnamefont {{Gioia}}},
  \bibinfo {author} {\bibfnamefont {G.}~\bibnamefont {{Luppino}}}, \bibinfo
  {author} {\bibfnamefont {J.~P.}\ \bibnamefont {{Hughes}}}, \ and\ \bibinfo
  {author} {\bibfnamefont {J.~T.}\ \bibnamefont {{Stocke}}},\ }\href {\doibase
  10.1086/305923} {\bibfield  {journal} {\bibinfo  {journal} {\apj}\ }\textbf
  {\bibinfo {volume} {502}},\ \bibinfo {pages} {550} (\bibinfo {year}
  {1998})},\ \Eprint {http://arxiv.org/abs/astro-ph/9707010}
  {arXiv:astro-ph/9707010 [astro-ph]} \BibitemShut {NoStop}%
\bibitem [{\citenamefont {{Hjorth}}\ \emph
  {et~al.}(1998{\natexlab{a}})\citenamefont {{Hjorth}}, \citenamefont
  {{Oukbir}},\ and\ \citenamefont {{van Kampen}}}]{Hjorth1998a}%
  \BibitemOpen
  \bibfield  {author} {\bibinfo {author} {\bibfnamefont {J.}~\bibnamefont
  {{Hjorth}}}, \bibinfo {author} {\bibfnamefont {J.}~\bibnamefont {{Oukbir}}},
  \ and\ \bibinfo {author} {\bibfnamefont {E.}~\bibnamefont {{van Kampen}}},\
  }\href {\doibase 10.1016/S1387-6473(98)00037-2} {\bibfield  {journal}
  {\bibinfo  {journal} {\nar}\ }\textbf {\bibinfo {volume} {42}},\ \bibinfo
  {pages} {145} (\bibinfo {year} {1998}{\natexlab{a}})}\BibitemShut {NoStop}%
\bibitem [{\citenamefont {{Hjorth}}\ \emph
  {et~al.}(1998{\natexlab{b}})\citenamefont {{Hjorth}}, \citenamefont
  {{Oukbir}},\ and\ \citenamefont {{van Kampen}}}]{Hjorth1998b}%
  \BibitemOpen
  \bibfield  {author} {\bibinfo {author} {\bibfnamefont {J.}~\bibnamefont
  {{Hjorth}}}, \bibinfo {author} {\bibfnamefont {J.}~\bibnamefont {{Oukbir}}},
  \ and\ \bibinfo {author} {\bibfnamefont {E.}~\bibnamefont {{van Kampen}}},\
  }\href {\doibase 10.1046/j.1365-8711.1998.01780.x} {\bibfield  {journal}
  {\bibinfo  {journal} {\mnras}\ }\textbf {\bibinfo {volume} {298}},\ \bibinfo
  {pages} {L1} (\bibinfo {year} {1998}{\natexlab{b}})},\ \Eprint
  {http://arxiv.org/abs/astro-ph/9802293} {arXiv:astro-ph/9802293 [astro-ph]}
  \BibitemShut {NoStop}%
\bibitem [{\citenamefont {{Mehrabi}}\ \emph {et~al.}(2017)\citenamefont
  {{Mehrabi}}, \citenamefont {{Pace}}, \citenamefont {{Malekjani}},\ and\
  \citenamefont {{Del Popolo}}}]{Mehrabi2017}%
  \BibitemOpen
  \bibfield  {author} {\bibinfo {author} {\bibfnamefont {A.}~\bibnamefont
  {{Mehrabi}}}, \bibinfo {author} {\bibfnamefont {F.}~\bibnamefont {{Pace}}},
  \bibinfo {author} {\bibfnamefont {M.}~\bibnamefont {{Malekjani}}}, \ and\
  \bibinfo {author} {\bibfnamefont {A.}~\bibnamefont {{Del Popolo}}},\ }\href
  {\doibase 10.1093/mnras/stw2927} {\bibfield  {journal} {\bibinfo  {journal}
  {\mnras}\ }\textbf {\bibinfo {volume} {465}},\ \bibinfo {pages} {2687}
  (\bibinfo {year} {2017})},\ \Eprint {http://arxiv.org/abs/1608.07961}
  {arXiv:1608.07961 [astro-ph.CO]} \BibitemShut {NoStop}%
\bibitem [{\citenamefont {{Cash}}(1979)}]{Cash1979}%
  \BibitemOpen
  \bibfield  {author} {\bibinfo {author} {\bibfnamefont {W.}~\bibnamefont
  {{Cash}}},\ }\href {\doibase 10.1086/156922} {\bibfield  {journal} {\bibinfo
  {journal} {\apj}\ }\textbf {\bibinfo {volume} {228}},\ \bibinfo {pages} {939}
  (\bibinfo {year} {1979})}\BibitemShut {NoStop}%
\bibitem [{\citenamefont {{Springel}}\ \emph {et~al.}(2005)\citenamefont
  {{Springel}}, \citenamefont {{White}}, \citenamefont {{Jenkins}},
  \citenamefont {{Frenk}}, \citenamefont {{Yoshida}}, \citenamefont {{Gao}},
  \citenamefont {{Navarro}}, \citenamefont {{Thacker}}, \citenamefont
  {{Croton}}, \citenamefont {{Helly}}, \citenamefont {{Peacock}}, \citenamefont
  {{Cole}}, \citenamefont {{Thomas}}, \citenamefont {{Couchman}}, \citenamefont
  {{Evrard}}, \citenamefont {{Colberg}},\ and\ \citenamefont
  {{Pearce}}}]{Springel2005a}%
  \BibitemOpen
  \bibfield  {author} {\bibinfo {author} {\bibfnamefont {V.}~\bibnamefont
  {{Springel}}}, \bibinfo {author} {\bibfnamefont {S.~D.~M.}\ \bibnamefont
  {{White}}}, \bibinfo {author} {\bibfnamefont {A.}~\bibnamefont {{Jenkins}}},
  \bibinfo {author} {\bibfnamefont {C.~S.}\ \bibnamefont {{Frenk}}}, \bibinfo
  {author} {\bibfnamefont {N.}~\bibnamefont {{Yoshida}}}, \bibinfo {author}
  {\bibfnamefont {L.}~\bibnamefont {{Gao}}}, \bibinfo {author} {\bibfnamefont
  {J.}~\bibnamefont {{Navarro}}}, \bibinfo {author} {\bibfnamefont
  {R.}~\bibnamefont {{Thacker}}}, \bibinfo {author} {\bibfnamefont
  {D.}~\bibnamefont {{Croton}}}, \bibinfo {author} {\bibfnamefont
  {J.}~\bibnamefont {{Helly}}}, \bibinfo {author} {\bibfnamefont {J.~A.}\
  \bibnamefont {{Peacock}}}, \bibinfo {author} {\bibfnamefont {S.}~\bibnamefont
  {{Cole}}}, \bibinfo {author} {\bibfnamefont {P.}~\bibnamefont {{Thomas}}},
  \bibinfo {author} {\bibfnamefont {H.}~\bibnamefont {{Couchman}}}, \bibinfo
  {author} {\bibfnamefont {A.}~\bibnamefont {{Evrard}}}, \bibinfo {author}
  {\bibfnamefont {J.}~\bibnamefont {{Colberg}}}, \ and\ \bibinfo {author}
  {\bibfnamefont {F.}~\bibnamefont {{Pearce}}},\ }\href {\doibase
  10.1038/nature03597} {\bibfield  {journal} {\bibinfo  {journal} {\nat}\
  }\textbf {\bibinfo {volume} {435}},\ \bibinfo {pages} {629} (\bibinfo {year}
  {2005})},\ \Eprint {http://arxiv.org/abs/astro-ph/0504097} {astro-ph/0504097}
  \BibitemShut {NoStop}%
\bibitem [{\citenamefont {{Peacock}}\ and\ \citenamefont
  {{Dodds}}(1996)}]{Peacock1996}%
  \BibitemOpen
  \bibfield  {author} {\bibinfo {author} {\bibfnamefont {J.~A.}\ \bibnamefont
  {{Peacock}}}\ and\ \bibinfo {author} {\bibfnamefont {S.~J.}\ \bibnamefont
  {{Dodds}}},\ }\href {http://adsabs.harvard.edu/abs/1996MNRAS.280L..19P}
  {\bibfield  {journal} {\bibinfo  {journal} {\mnras}\ }\textbf {\bibinfo
  {volume} {280}},\ \bibinfo {pages} {L19} (\bibinfo {year} {1996})},\ \Eprint
  {http://arxiv.org/abs/arXiv:astro-ph/9603031} {arXiv:astro-ph/9603031}
  \BibitemShut {NoStop}%
\bibitem [{\citenamefont {{Zhao}}(2014)}]{Zhao2014}%
  \BibitemOpen
  \bibfield  {author} {\bibinfo {author} {\bibfnamefont {G.-B.}\ \bibnamefont
  {{Zhao}}},\ }\href {\doibase 10.1088/0067-0049/211/2/23} {\bibfield
  {journal} {\bibinfo  {journal} {\apjs}\ }\textbf {\bibinfo {volume} {211}},\
  \bibinfo {eid} {23} (\bibinfo {year} {2014})},\ \Eprint
  {http://arxiv.org/abs/1312.1291} {arXiv:1312.1291} \BibitemShut {NoStop}%
\bibitem [{\citenamefont {{Schneider}}\ \emph {et~al.}(2012)\citenamefont
  {{Schneider}}, \citenamefont {{Smith}}, \citenamefont {{Macci{\`o}}},\ and\
  \citenamefont {{Moore}}}]{Schneider2012}%
  \BibitemOpen
  \bibfield  {author} {\bibinfo {author} {\bibfnamefont {A.}~\bibnamefont
  {{Schneider}}}, \bibinfo {author} {\bibfnamefont {R.~E.}\ \bibnamefont
  {{Smith}}}, \bibinfo {author} {\bibfnamefont {A.~V.}\ \bibnamefont
  {{Macci{\`o}}}}, \ and\ \bibinfo {author} {\bibfnamefont {B.}~\bibnamefont
  {{Moore}}},\ }\href {\doibase 10.1111/j.1365-2966.2012.21252.x} {\bibfield
  {journal} {\bibinfo  {journal} {\mnras}\ }\textbf {\bibinfo {volume} {424}},\
  \bibinfo {pages} {684} (\bibinfo {year} {2012})},\ \Eprint
  {http://arxiv.org/abs/1112.0330} {arXiv:1112.0330 [astro-ph.CO]} \BibitemShut
  {NoStop}%
\bibitem [{\citenamefont {{Marsh}}(2016)}]{Marsh2016}%
  \BibitemOpen
  \bibfield  {author} {\bibinfo {author} {\bibfnamefont {D.~J.~E.}\
  \bibnamefont {{Marsh}}},\ }\href@noop {} {\bibfield  {journal} {\bibinfo
  {journal} {arXiv e-prints}\ ,\ \bibinfo {eid} {arXiv:1605.05973}} (\bibinfo
  {year} {2016})},\ \Eprint {http://arxiv.org/abs/1605.05973} {arXiv:1605.05973
  [astro-ph.CO]} \BibitemShut {NoStop}%
\bibitem [{\citenamefont {{Takahashi}}\ \emph {et~al.}(2012)\citenamefont
  {{Takahashi}}, \citenamefont {{Sato}}, \citenamefont {{Nishimichi}},
  \citenamefont {{Taruya}},\ and\ \citenamefont {{Oguri}}}]{Takahashi2012}%
  \BibitemOpen
  \bibfield  {author} {\bibinfo {author} {\bibfnamefont {R.}~\bibnamefont
  {{Takahashi}}}, \bibinfo {author} {\bibfnamefont {M.}~\bibnamefont {{Sato}}},
  \bibinfo {author} {\bibfnamefont {T.}~\bibnamefont {{Nishimichi}}}, \bibinfo
  {author} {\bibfnamefont {A.}~\bibnamefont {{Taruya}}}, \ and\ \bibinfo
  {author} {\bibfnamefont {M.}~\bibnamefont {{Oguri}}},\ }\href {\doibase
  10.1088/0004-637X/761/2/152} {\bibfield  {journal} {\bibinfo  {journal}
  {\apj}\ }\textbf {\bibinfo {volume} {761}},\ \bibinfo {eid} {152} (\bibinfo
  {year} {2012})},\ \Eprint {http://arxiv.org/abs/1208.2701} {arXiv:1208.2701
  [astro-ph.CO]} \BibitemShut {NoStop}%
\bibitem [{\citenamefont {{Spurio Mancini}}\ \emph {et~al.}(2019)\citenamefont
  {{Spurio Mancini}}, \citenamefont {{K{\"o}hlinger}}, \citenamefont
  {{Joachimi}}, \citenamefont {{Pettorino}}, \citenamefont {{Sch{\"a}fer}},
  \citenamefont {{Reischke}}, \citenamefont {{van Uitert}}, \citenamefont
  {{Brieden}}, \citenamefont {{Archidiacono}},\ and\ \citenamefont
  {{Lesgourgues}}}]{SpurioMancini2019}%
  \BibitemOpen
  \bibfield  {author} {\bibinfo {author} {\bibfnamefont {A.}~\bibnamefont
  {{Spurio Mancini}}}, \bibinfo {author} {\bibfnamefont {F.}~\bibnamefont
  {{K{\"o}hlinger}}}, \bibinfo {author} {\bibfnamefont {B.}~\bibnamefont
  {{Joachimi}}}, \bibinfo {author} {\bibfnamefont {V.}~\bibnamefont
  {{Pettorino}}}, \bibinfo {author} {\bibfnamefont {B.~M.}\ \bibnamefont
  {{Sch{\"a}fer}}}, \bibinfo {author} {\bibfnamefont {R.}~\bibnamefont
  {{Reischke}}}, \bibinfo {author} {\bibfnamefont {E.}~\bibnamefont {{van
  Uitert}}}, \bibinfo {author} {\bibfnamefont {S.}~\bibnamefont {{Brieden}}},
  \bibinfo {author} {\bibfnamefont {M.}~\bibnamefont {{Archidiacono}}}, \ and\
  \bibinfo {author} {\bibfnamefont {J.}~\bibnamefont {{Lesgourgues}}},\ }\href
  {\doibase 10.1093/mnras/stz2581} {\bibfield  {journal} {\bibinfo  {journal}
  {\mnras}\ }\textbf {\bibinfo {volume} {490}},\ \bibinfo {pages} {2155}
  (\bibinfo {year} {2019})},\ \Eprint {http://arxiv.org/abs/1901.03686}
  {arXiv:1901.03686 [astro-ph.CO]} \BibitemShut {NoStop}%
\bibitem [{\citenamefont {{Hildebrandt}}\ \emph {et~al.}(2017)\citenamefont
  {{Hildebrandt}}, \citenamefont {{Viola}}, \citenamefont {{Heymans}},
  \citenamefont {{Joudaki}}, \citenamefont {{Kuijken}}, \citenamefont
  {{Blake}}, \citenamefont {{Erben}}, \citenamefont {{Joachimi}}, \citenamefont
  {{Klaes}},\ and\ \citenamefont {{et~al.}}}]{Hildebrandt2017}%
  \BibitemOpen
  \bibfield  {author} {\bibinfo {author} {\bibfnamefont {H.}~\bibnamefont
  {{Hildebrandt}}}, \bibinfo {author} {\bibfnamefont {M.}~\bibnamefont
  {{Viola}}}, \bibinfo {author} {\bibfnamefont {C.}~\bibnamefont {{Heymans}}},
  \bibinfo {author} {\bibfnamefont {S.}~\bibnamefont {{Joudaki}}}, \bibinfo
  {author} {\bibfnamefont {K.}~\bibnamefont {{Kuijken}}}, \bibinfo {author}
  {\bibfnamefont {C.}~\bibnamefont {{Blake}}}, \bibinfo {author} {\bibfnamefont
  {T.}~\bibnamefont {{Erben}}}, \bibinfo {author} {\bibfnamefont
  {B.}~\bibnamefont {{Joachimi}}}, \bibinfo {author} {\bibfnamefont
  {D.}~\bibnamefont {{Klaes}}}, \ and\ \bibinfo {author} {\bibnamefont
  {{et~al.}}},\ }\href {\doibase 10.1093/mnras/stw2805} {\bibfield  {journal}
  {\bibinfo  {journal} {\mnras}\ }\textbf {\bibinfo {volume} {465}},\ \bibinfo
  {pages} {1454} (\bibinfo {year} {2017})},\ \Eprint
  {http://arxiv.org/abs/1606.05338} {arXiv:1606.05338 [astro-ph.CO]}
  \BibitemShut {NoStop}%
\bibitem [{\citenamefont {{K{\"o}hlinger}}\ \emph {et~al.}(2017)\citenamefont
  {{K{\"o}hlinger}}, \citenamefont {{Viola}}, \citenamefont {{Joachimi}},
  \citenamefont {{Hoekstra}}, \citenamefont {{van Uitert}}, \citenamefont
  {{Hildebrandt}}, \citenamefont {{Choi}}, \citenamefont {{Erben}},
  \citenamefont {{Heymans}},\ and\ \citenamefont {{et~al.}}}]{Kohlinger2017}%
  \BibitemOpen
  \bibfield  {author} {\bibinfo {author} {\bibfnamefont {F.}~\bibnamefont
  {{K{\"o}hlinger}}}, \bibinfo {author} {\bibfnamefont {M.}~\bibnamefont
  {{Viola}}}, \bibinfo {author} {\bibfnamefont {B.}~\bibnamefont {{Joachimi}}},
  \bibinfo {author} {\bibfnamefont {H.}~\bibnamefont {{Hoekstra}}}, \bibinfo
  {author} {\bibfnamefont {E.}~\bibnamefont {{van Uitert}}}, \bibinfo {author}
  {\bibfnamefont {H.}~\bibnamefont {{Hildebrandt}}}, \bibinfo {author}
  {\bibfnamefont {A.}~\bibnamefont {{Choi}}}, \bibinfo {author} {\bibfnamefont
  {T.}~\bibnamefont {{Erben}}}, \bibinfo {author} {\bibfnamefont
  {C.}~\bibnamefont {{Heymans}}}, \ and\ \bibinfo {author} {\bibnamefont
  {{et~al.}}},\ }\href {\doibase 10.1093/mnras/stx1820} {\bibfield  {journal}
  {\bibinfo  {journal} {\mnras}\ }\textbf {\bibinfo {volume} {471}},\ \bibinfo
  {pages} {4412} (\bibinfo {year} {2017})},\ \Eprint
  {http://arxiv.org/abs/1706.02892} {arXiv:1706.02892 [astro-ph.CO]}
  \BibitemShut {NoStop}%
\bibitem [{\citenamefont {{Driver}}\ \emph {et~al.}(2009)\citenamefont
  {{Driver}}, \citenamefont {{Norberg}}, \citenamefont {{Baldry}},
  \citenamefont {{Bamford}}, \citenamefont {{Hopkins}}, \citenamefont
  {{Liske}}, \citenamefont {{Loveday}}, \citenamefont {{Peacock}},
  \citenamefont {{Hill}},\ and\ \citenamefont {{et~al.}}}]{Driver2009}%
  \BibitemOpen
  \bibfield  {author} {\bibinfo {author} {\bibfnamefont {S.~P.}\ \bibnamefont
  {{Driver}}}, \bibinfo {author} {\bibfnamefont {P.}~\bibnamefont {{Norberg}}},
  \bibinfo {author} {\bibfnamefont {I.~K.}\ \bibnamefont {{Baldry}}}, \bibinfo
  {author} {\bibfnamefont {S.~P.}\ \bibnamefont {{Bamford}}}, \bibinfo {author}
  {\bibfnamefont {A.~M.}\ \bibnamefont {{Hopkins}}}, \bibinfo {author}
  {\bibfnamefont {J.}~\bibnamefont {{Liske}}}, \bibinfo {author} {\bibfnamefont
  {J.}~\bibnamefont {{Loveday}}}, \bibinfo {author} {\bibfnamefont {J.~A.}\
  \bibnamefont {{Peacock}}}, \bibinfo {author} {\bibfnamefont {D.~T.}\
  \bibnamefont {{Hill}}}, \ and\ \bibinfo {author} {\bibnamefont {{et~al.}}},\
  }\href {\doibase 10.1111/j.1468-4004.2009.50512.x} {\bibfield  {journal}
  {\bibinfo  {journal} {Astronomy and Geophysics}\ }\textbf {\bibinfo {volume}
  {50}},\ \bibinfo {pages} {5.12} (\bibinfo {year} {2009})},\ \Eprint
  {http://arxiv.org/abs/0910.5123} {arXiv:0910.5123 [astro-ph.CO]} \BibitemShut
  {NoStop}%
\bibitem [{\citenamefont {{Driver}}\ \emph {et~al.}(2011)\citenamefont
  {{Driver}}, \citenamefont {{Hill}}, \citenamefont {{Kelvin}}, \citenamefont
  {{Robotham}}, \citenamefont {{Liske}}, \citenamefont {{Norberg}},
  \citenamefont {{Baldry}}, \citenamefont {{Bamford}}, \citenamefont
  {{Hopkins}},\ and\ \citenamefont {{et~al.}}}]{Driver2011}%
  \BibitemOpen
  \bibfield  {author} {\bibinfo {author} {\bibfnamefont {S.~P.}\ \bibnamefont
  {{Driver}}}, \bibinfo {author} {\bibfnamefont {D.~T.}\ \bibnamefont
  {{Hill}}}, \bibinfo {author} {\bibfnamefont {L.~S.}\ \bibnamefont
  {{Kelvin}}}, \bibinfo {author} {\bibfnamefont {A.~S.~G.}\ \bibnamefont
  {{Robotham}}}, \bibinfo {author} {\bibfnamefont {J.}~\bibnamefont {{Liske}}},
  \bibinfo {author} {\bibfnamefont {P.}~\bibnamefont {{Norberg}}}, \bibinfo
  {author} {\bibfnamefont {I.~K.}\ \bibnamefont {{Baldry}}}, \bibinfo {author}
  {\bibfnamefont {S.~P.}\ \bibnamefont {{Bamford}}}, \bibinfo {author}
  {\bibfnamefont {A.~M.}\ \bibnamefont {{Hopkins}}}, \ and\ \bibinfo {author}
  {\bibnamefont {{et~al.}}},\ }\href {\doibase
  10.1111/j.1365-2966.2010.18188.x} {\bibfield  {journal} {\bibinfo  {journal}
  {\mnras}\ }\textbf {\bibinfo {volume} {413}},\ \bibinfo {pages} {971}
  (\bibinfo {year} {2011})},\ \Eprint {http://arxiv.org/abs/1009.0614}
  {arXiv:1009.0614 [astro-ph.CO]} \BibitemShut {NoStop}%
\bibitem [{\citenamefont {{Liske}}\ \emph {et~al.}(2015)\citenamefont
  {{Liske}}, \citenamefont {{Baldry}}, \citenamefont {{Driver}}, \citenamefont
  {{Tuffs}}, \citenamefont {{Alpaslan}}, \citenamefont {{Andrae}},
  \citenamefont {{Brough}}, \citenamefont {{Cluver}}, \citenamefont
  {{Grootes}},\ and\ \citenamefont {{et~al.}}}]{Liske2015}%
  \BibitemOpen
  \bibfield  {author} {\bibinfo {author} {\bibfnamefont {J.}~\bibnamefont
  {{Liske}}}, \bibinfo {author} {\bibfnamefont {I.~K.}\ \bibnamefont
  {{Baldry}}}, \bibinfo {author} {\bibfnamefont {S.~P.}\ \bibnamefont
  {{Driver}}}, \bibinfo {author} {\bibfnamefont {R.~J.}\ \bibnamefont
  {{Tuffs}}}, \bibinfo {author} {\bibfnamefont {M.}~\bibnamefont {{Alpaslan}}},
  \bibinfo {author} {\bibfnamefont {E.}~\bibnamefont {{Andrae}}}, \bibinfo
  {author} {\bibfnamefont {S.}~\bibnamefont {{Brough}}}, \bibinfo {author}
  {\bibfnamefont {M.~E.}\ \bibnamefont {{Cluver}}}, \bibinfo {author}
  {\bibfnamefont {M.~W.}\ \bibnamefont {{Grootes}}}, \ and\ \bibinfo {author}
  {\bibnamefont {{et~al.}}},\ }\href {\doibase 10.1093/mnras/stv1436}
  {\bibfield  {journal} {\bibinfo  {journal} {\mnras}\ }\textbf {\bibinfo
  {volume} {452}},\ \bibinfo {pages} {2087} (\bibinfo {year} {2015})},\ \Eprint
  {http://arxiv.org/abs/1506.08222} {arXiv:1506.08222 [astro-ph.GA]}
  \BibitemShut {NoStop}%
\bibitem [{\citenamefont {{Ili{\'c}}}\ \emph {et~al.}(2019)\citenamefont
  {{Ili{\'c}}}, \citenamefont {{Sakr}},\ and\ \citenamefont
  {{Blanchard}}}]{Ilic2019}%
  \BibitemOpen
  \bibfield  {author} {\bibinfo {author} {\bibfnamefont {S.}~\bibnamefont
  {{Ili{\'c}}}}, \bibinfo {author} {\bibfnamefont {Z.}~\bibnamefont {{Sakr}}},
  \ and\ \bibinfo {author} {\bibfnamefont {A.}~\bibnamefont {{Blanchard}}},\
  }\href {\doibase 10.1051/0004-6361/201936423} {\bibfield  {journal} {\bibinfo
   {journal} {\aap}\ }\textbf {\bibinfo {volume} {631}},\ \bibinfo {eid} {A96}
  (\bibinfo {year} {2019})}\BibitemShut {NoStop}%
\end{thebibliography}%

\label{lastpage}

\end{document}